\documentclass[12pt,letterpaper]{article}
\usepackage[a4paper, total={7in, 10in}]{geometry}

\usepackage{graphicx}
\usepackage{helvet}
\usepackage{authblk}
\usepackage{hyperref}
\hypersetup{hidelinks}
\usepackage{amsmath}
\usepackage{amssymb}
\usepackage{orcidlink}
\usepackage{booktabs}
\usepackage[square, numbers, sort&compress]{natbib}
\usepackage[right]{lineno}
\usepackage{eurosym}
\usepackage{setspace}
\usepackage{longtable}
\usepackage{caption}
\usepackage{titlesec}
\usepackage{tabularx}
\makeatletter
\renewcommand{\maketitle}{\bgroup\setlength{\parindent}{0pt}
\begin{flushleft}
  \textbf{\@title}

  \@author
\end{flushleft}\egroup}
\makeatother

\title{Why delivered costs understate prices in an integrated global hydrogen market}
\date{}

\author[1,2,*,\orcidlink{0009-0003-3018-6811}]{B. Wei\ss{}enburger}
\author[1]{M. Wietschel}
\author[1]{P. Pl\"otz}
\author[2,3]{R. McKenna}

\affil[1]{Fraunhofer Institute for Systems and Innovation Research ISI, Breslauer Stra\ss e 48, 76139 Karlsruhe, Germany}
\affil[2]{Chair of Energy Systems Analysis, Institute of Energy and Process Engineering, ETH Z\"urich, Switzerland}
\affil[3]{Laboratory for Energy Systems Analysis, Paul Scherrer Institute, Villigen, Switzerland}
\affil[*]{Corresponding author: bastian.weissenburger@isi.fraunhofer.de}

\begin{document}

\maketitle

\section*{Abstract}

Delivered-cost estimates are often used as proxies for hydrogen import prices. Yet they measure average supply-chain costs, whereas marginal prices measure the system value of an additional unit. Market interactions can make the two diverge, potentially biasing long-term price assumptions. We quantify this gap using an 81-region global hydrogen-market model, comparing each region’s costs and prices when only its demand is active and when all regions’ demands are active. The 2045 price--cost gap is positive in 78 regions and averages 63 EUR/MWh (52\% of the single-buyer cost) on a demand-weighted basis. Only 12 EUR/MWh reflects higher average sourcing costs due to competing demand. The remaining 51 EUR/MWh is the difference between the global-market price and the delivered portfolio's average cost. In several regions, wholesale prices rise while average delivered costs fall. Hydrogen strategies should distinguish delivered-cost benchmarks from market-price expectations, as low-cost supply does not guarantee low regional market prices.

\section*{Introduction}

Many governments see low-carbon hydrogen as an important option for defossilising hard-to-electrify sectors, and national hydrogen strategies often anticipate substantial imports from regions with low production costs \cite{GermanFederalMinistryforEconomicAffairsandClimateAction.2023b,EuropeanCommission.2022,GermanFederalMinistryforEconomicAffairsandClimateAction.2024, MinisterialCouncilonRenewableEnergyofJapan.}. Progress has nevertheless fallen short of announced ambitions \cite{InternationalEnergyAgency.2025,Odenweller.2025,Terlouw.2025}. Uncertainty about future demand and prices remains an important barrier to long-term offtake agreements and investment \cite{InternationalEnergyAgency.2025,Wietschel.2021}. Because projected hydrogen demand responds strongly to assumed prices, expectations of future prices are central to these decisions \cite{Marmullaku.2026,Weienburger.2024}. Yet few studies estimate future regional market prices \cite{Schonfisch.2022,Alanazi.2025,Alanazi.2026,Barner.2024,Antweiler.2024}. Many published studies instead report production or delivered costs \cite{Moritz.2023,Genge.2023,Hampp.2023,Brandle.2021}, and these estimates vary widely. For Europe in 2050, they differ by approximately a factor of five \cite{Genge.2023}. The practical question is therefore not only where hydrogen can be produced and delivered at low cost, but also what price a region may face for the volumes it requires.

These delivered-cost estimates generally come from route-level supply-chain studies \cite{Moritz.2023,Genge.2023,Hampp.2023,Brandle.2021}. They typically combine the levelised cost of producing hydrogen at a specified location with the costs of conversion, storage and transport along a particular route, at an assumed scale and utilisation rate. By themselves, however, these estimates do not show what price may emerge when several suppliers and buyers interact. Cost-derived assumptions are nevertheless used as regional import-price inputs in energy-system models that do not represent market prices endogenously \cite{Keutz.2025,Schob.2023, Moskalenko.2026,Fleiter.2024}, and some national strategies use supply-cost targets as planning benchmarks \cite{MinisterialCouncilonRenewableEnergyofJapan.}.

Using delivered costs as price proxies may be reasonable when low-cost production and transport remain plentiful at the required volume, so that average delivered cost remains close to the marginal value of supplying the region. Global assessments suggest that low-cost production potential exceeds projected demand by a wide margin \cite{Shirizadeh.2023}, although much of it lies away from demand centres and is constrained by water, land and renewable electricity availability \cite{Terlouw.2024}. Production potential available at a given cost is not the same as supply delivered to a region at that cost, once transport and demand elsewhere are taken into account. A region facing no competing demand from other regions uses the cheapest supply options first and turns to more expensive ones as demand increases. The average cost of its supply portfolio can therefore remain below the marginal value of supplying an additional unit, creating a price--cost gap even when the region is the only buyer. When other regions also demand hydrogen, they draw on the same finite production resources and transport routes. This can change both the average cost of the supply reaching a region and its marginal price. A region may even receive a cheaper average portfolio while facing a higher price.

The supply-chain studies discussed above represent demand only for the region they investigate, or not at all \cite{Fakhreddine.2025}. Integrated hydrogen-market models instead determine production, trade and infrastructure jointly. Many report traded volumes, routes and delivered costs without calculating regional prices \cite{IRENA.2022,NunezJimenez.2022b,Linsel.2023}. Where regional prices are derived, they are used to interpret outcomes across scenarios, market structure or trade regimes \cite{Schonfisch.2022,Alanazi.2025,Barner.2024,Alanazi.2026,Antweiler.2024}. To our knowledge, these prices have not been compared against the delivered cost of the supply reaching each region. We therefore ask how large the resulting price--cost gap is, how competing demand changes prices, and whether the gap mainly reflects a change in sourcing or prices above average delivered costs.

\begin{figure}[!htbp]
    \centering
    \includegraphics{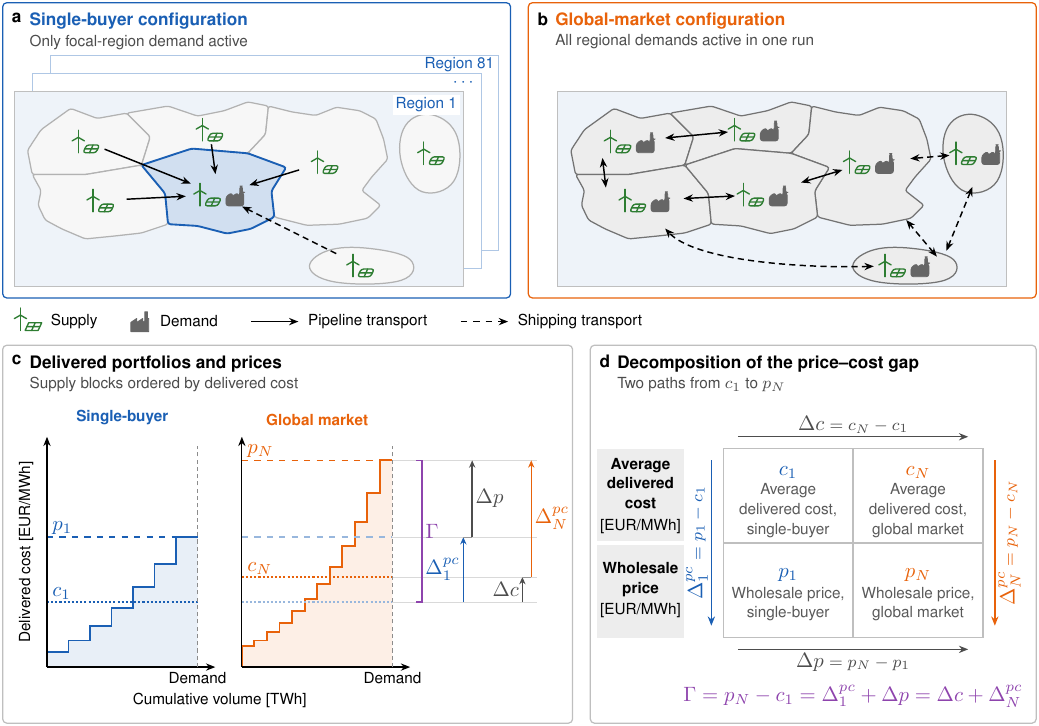}
    \caption{\textbf{Matched model configurations and decomposition of the hydrogen price--cost gap.} (\textbf{a}) In the single-buyer configuration, only the focal region has active hydrogen demand, while the full set of global production resources and transport options remains available. Repeating the model for each focal region produces 81 matched single-buyer runs. (\textbf{b}) In the global-market configuration, all 81 regions have active demand and can produce and trade hydrogen in a single run. The focal region’s demand and all techno-economic assumptions are unchanged between configurations, while investment, production and trade respond endogenously. (\textbf{c}) Schematic cost-ordered supply blocks show the hydrogen portfolio delivered to the same focal-region demand in each configuration. The quantities $c_1$ and $c_N$ are the volume-weighted average delivered costs in the single-buyer and global-market configurations, respectively. The quantities $p_1$ and $p_N$ are the dual values of the regional hydrogen-balance constraint in the fixed-capacity pricing stage and are reported as modelled wholesale prices. These prices reflect marginal system values and should not necessarily be interpreted as the cost of the most expensive displayed supply block. (\textbf{d}) The price--cost gap, $\Gamma=p_N-c_1$, can be decomposed along two paths. The first gives $\Gamma=\Delta^{pc}_{1}+\Delta p$, where $\Delta^{pc}_{1}=p_1-c_1$ is the single-buyer price--cost difference and $\Delta p=p_N-p_1$ is the price effect of competing demand. The second gives $\Gamma=\Delta c+\Delta^{pc}_{N}$, where $\Delta c=c_N-c_1$ is the sourcing effect of competing demand and $\Delta^{pc}_{N}=p_N-c_N$ is the global-market price--cost difference. The maps and supply blocks are schematic and contain no numerical results. For readability, year and region subscripts are omitted from the notation in the main text and this figure. The Supplementary Information uses the full notation, for example $(p_{N,y,i})$.}
    \label{fig:CompetitionPremium}
\end{figure}

We address these questions using HyPriM \cite{FraunhoferInstituteforSystemsandInnovationResearchISI.2026}, a global hydrogen-market optimisation model covering 81 regions from 2030 to 2050, building on \cite{Wietschel.2024b}. For each focal region, we compare two matched configurations. In the single-buyer configuration, only the focal region's prescribed demand is active, but all modelled production potentials and transport options remain available. This extends the cost logic of route studies to the region’s full demand, rather than replicating an individual study. In the global-market configuration, all regions have active demand and can produce and trade simultaneously. The focal region’s demand and techno-economic assumptions are unchanged, while investment, production and trade respond endogenously. In both configurations, we calculate the volume-weighted average delivered cost and the dual value of the regional hydrogen balance in the fixed-capacity pricing stage. We report the latter as a modelled long-run wholesale price conditional on the installed capacities and price-taking behaviour \cite{ONeill.2005,Byers.2023}. The resulting four quantities allow two decompositions of the price--cost gap (Fig.~\ref{fig:CompetitionPremium}). One separates the single-buyer price–cost difference from the price effect of competing demand. The other separates the sourcing effect, the change in average delivered cost, from the global-market price–cost difference. We examine supply changes for Germany and Japan in more detail. Germany has many pipeline options within Europe, whereas Japan has few pipeline options and can import from many potential suppliers only by ship. We find that modelled global-market prices exceed the matched cost benchmark in nearly all regions. On a demand-weighted basis, most of this difference reflects prices above average delivered costs rather than a more expensive supply portfolio. Delivered costs therefore remain useful for comparing supply options but should not be used directly as long-term price assumptions in hydrogen strategies and energy-system models.

\section*{Results}

\subsection*{Market prices exceed supply-chain cost benchmarks in nearly all regions}

\begin{figure}[!htbp]
    \centering
    \includegraphics[width=\textwidth]{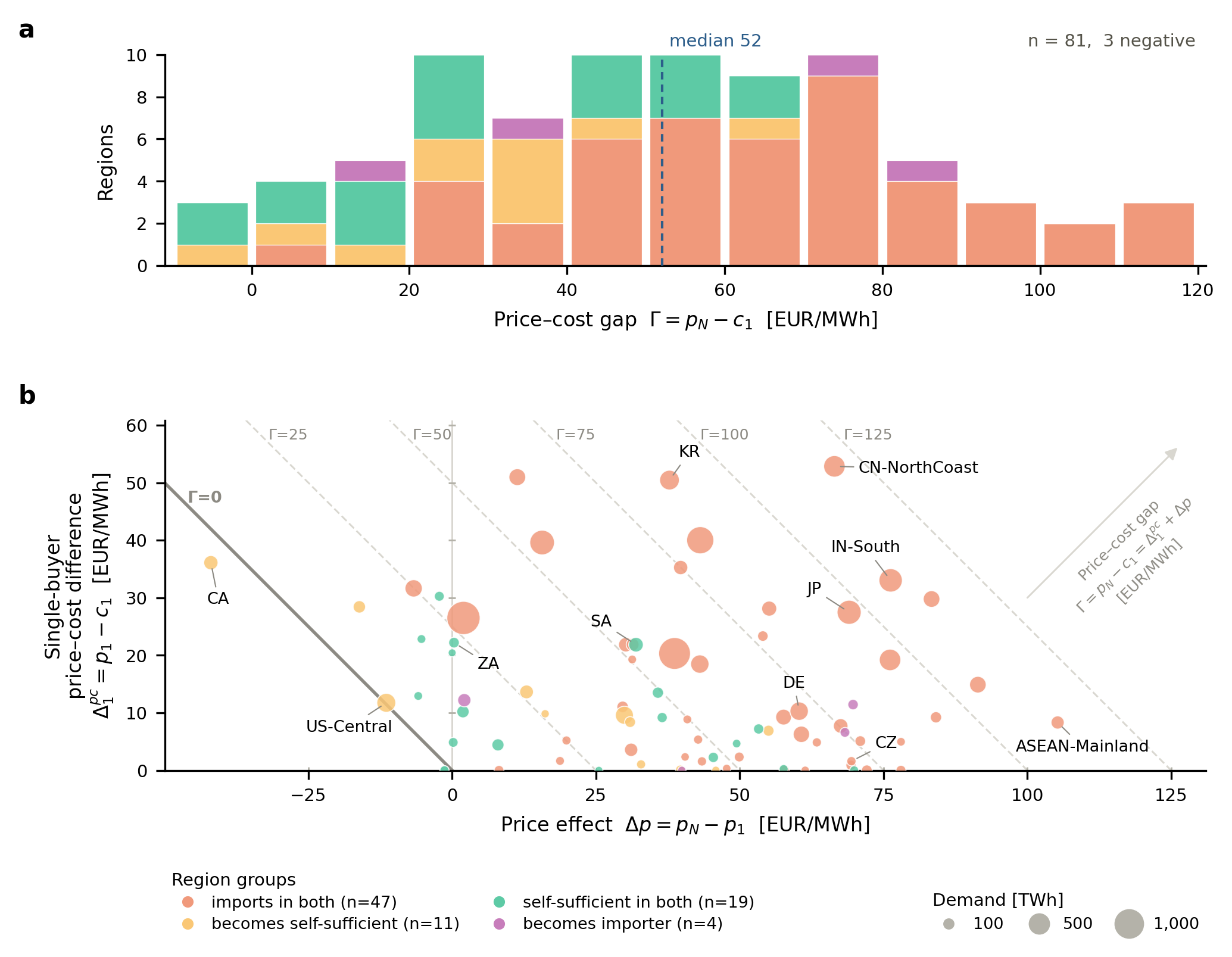}
    \caption{\textbf{Regional decomposition of the hydrogen price--cost gap in 2045.}
    (a) Distribution of the price--cost gap, $\Gamma$, across all 81 regions. Bar heights indicate the number of regions, with colours distinguishing their import status.
    (b) The price effect of competing demand, $\Delta p=p_N-p_1$, and the single-buyer price--cost difference, $\Delta^{pc}_{1}=p_1-c_1$, are shown for the same regions. Here, $c_1$ and $p_1$ are the average delivered cost and wholesale price when only the focal region has active hydrogen demand, and $p_N$ is the wholesale price when all regions have active demand. Together, the two components give the overall price--cost gap, $\Gamma=\Delta^{pc}_{1}+\Delta p=p_N-c_1$. Grey diagonal lines indicate constant values of $\Gamma$. Marker area is proportional to regional hydrogen demand. Colours in both panels distinguish regions that import in both configurations, remain self-sufficient in both, become self-sufficient or become importers when moving from the single-buyer to the global-market configuration. Import status refers to whether any of the hydrogen consumed in a region is produced elsewhere, irrespective of its exports.}
    \label{fig:gap}
\end{figure}

Figure~\ref{fig:gap}a shows the price--cost gap, $\Gamma$, across all 81 regions in 2045. Of these regions, 78 show a positive gap. Across all regions, the median gap is 52~EUR/MWh (interquartile range of 28--71~EUR/MWh) and the median relative gap is 45\% of the single-buyer cost benchmark. On a demand-weighted basis, the global-market price is 183~EUR/MWh and the single-buyer delivered-cost benchmark is 120~EUR/MWh, resulting in a gap of 63~EUR/MWh, or 52\%.

Positive gaps occur in all four region groups, although their magnitude differs. Regions that import in both configurations have the highest median gap of 67~EUR/MWh, whereas regions that are self-sufficient in both configurations have the lowest, at 26~EUR/MWh. The latter group nonetheless includes large gaps, reaching 70~EUR/MWh. Only three regions have a negative total gap in 2045, all self-sufficient in the global-market configuration. Canada is the only sizeable case, at $-6$~EUR/MWh. Across 2035--2050, the median regional gap remains between 51 and 69~EUR/MWh, so the gap also persists along the pathway.

The relative contributions of the single-buyer price--cost difference, $\Delta^{pc}_{1}$, and the price effect of competing demand, $\Delta p$, vary widely across regions (Fig.~\ref{fig:gap}b). On a demand-weighted basis, $\Delta^{pc}_{1}$ amounts to 24~EUR/MWh, while activating demand in the other regions raises the price by approximately 39~EUR/MWh. Part of the gap therefore already exists before any competing buyer is active, and part is attributable to the price effect of competing demand. In 16 of the 78 positive-gap regions, accounting for 30\% of demand in that group, the single-buyer price–cost difference exceeds the price effect of competing demand. In Korea, for example, these contributions are 50 and 38~EUR/MWh, respectively. By contrast, in the Czech Republic the price effect is 69 EUR/MWh and the single-buyer difference only 2 EUR/MWh. The price effect is negative in nine regions, eight of which are self-sufficient in the global-market configuration. The overall price--cost gap therefore cannot be attributed to competing demand alone, and competing demand does not necessarily raise the price in every region.

\subsection*{Competing demand reallocates low-cost supply among buyers}

\begin{figure}[htbp]
    \centering
    \includegraphics[width=\textwidth]{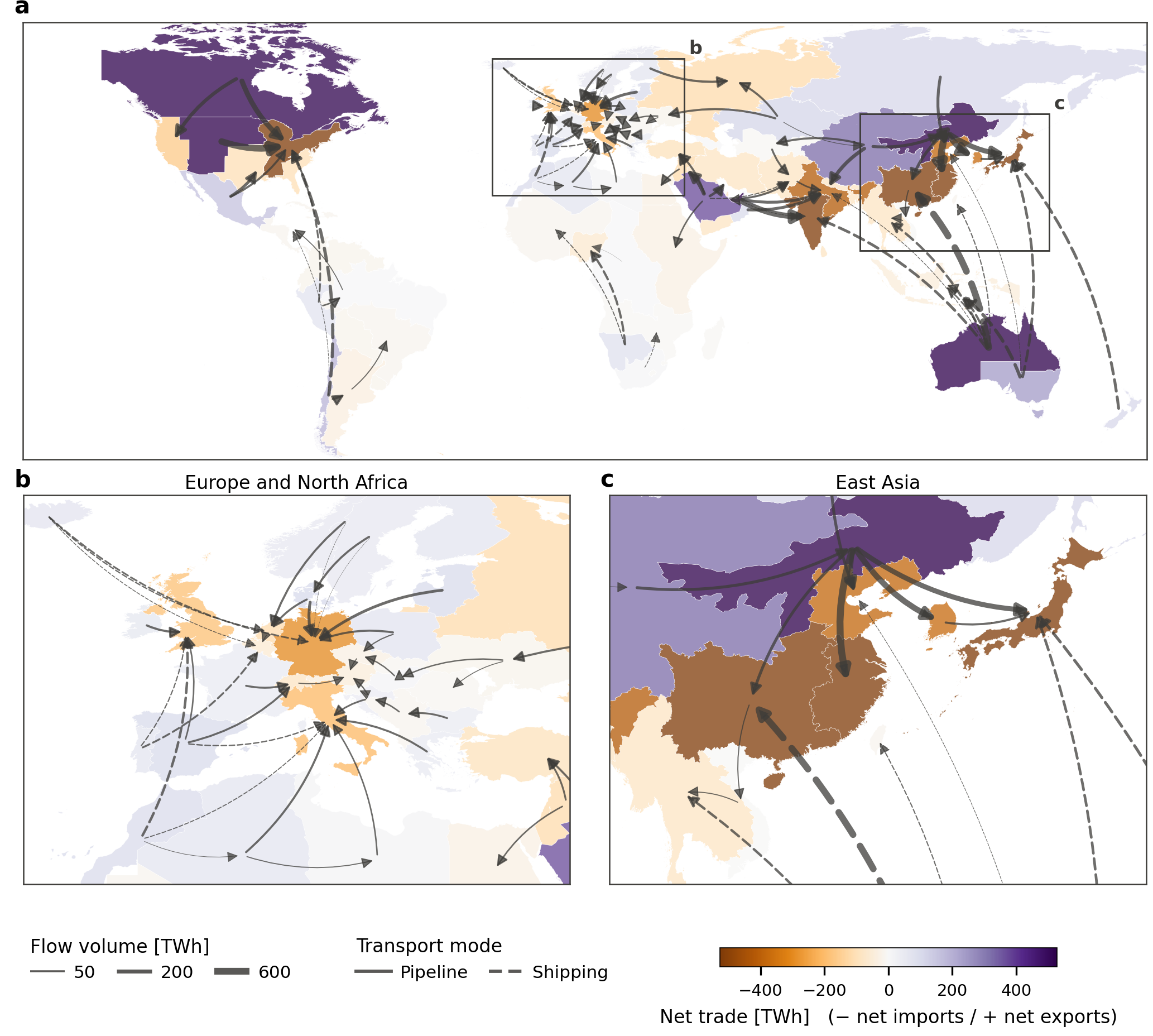}
    \caption{\textbf{International hydrogen trade in the global-market configuration in 2045.} (a) Global trade network. (b,c) Zoomed-in views of Europe and North Africa (b) and East Asia (c). Lines show physical hydrogen flows between regions. Line width is proportional to annual corridor flow, arrows indicate the direction of flow, and line style distinguishes pipeline from shipping transport. Region colour shows the net trade balance, with positive values indicating net exports and negative values indicating net imports. In total, 7,154~TWh of hydrogen is consumed outside its production region, equivalent to 57\% of demand across all 81 regions. This involves 43 exporting and 51 importing regions over 99 active corridors, comprising 68 pipeline and 31 shipping corridors. A region can both import and export hydrogen, so the importer and exporter groups are not mutually exclusive.}
    \label{fig:trade}
\end{figure}

The price--cost gap arises in an interconnected international market rather than in isolated regions (Fig.~\ref{fig:trade}). In 2045, 7,154~TWh of hydrogen is consumed outside its production region, equivalent to 57\% of global demand. The structure of this trade differs across world regions. East Asia has several large importers and a few large exporters. European regions generally have smaller import or export balances and are connected through a denser regional network. We examine how activating demand elsewhere reshapes supply for Germany and Japan, which differ structurally in their supply options.

Germany and Japan receive the same hydrogen volumes in both configurations (318~TWh and 595~TWh, respectively), but their supply portfolios change. In both cases, major suppliers in the single-buyer configuration increase their total production when demand elsewhere is activated, but allocate a smaller share to the focal country.

\begin{figure}[p]
    \centering
    \includegraphics[width=\textwidth]{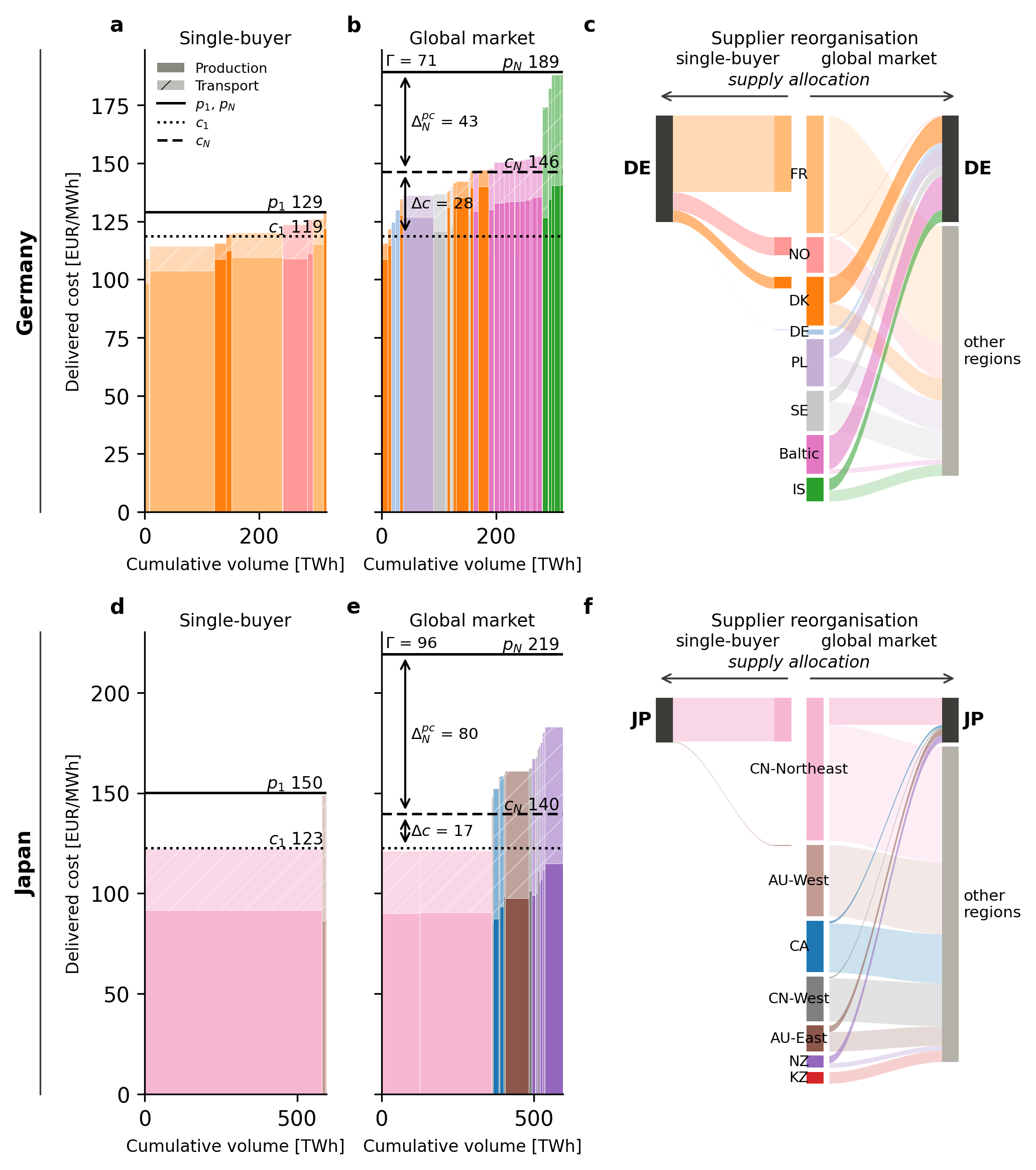}
    \caption{\textbf{Hydrogen supply portfolios of Germany and Japan for the single-buyer and global-market configurations in 2045.} \textbf{a,b}, Supply portfolios serving Germany in 2045 in the single-buyer (\textbf{a}) and global-market (\textbf{b}) configurations. Only the focal country has active demand in the single-buyer configuration, all modelled regions have active demand in the global-market configuration. Supply blocks are ordered by delivered cost, and their widths show volumes delivered after transport losses. Each block is divided into production and transport costs. Horizontal lines show the wholesale price and volume-weighted average delivered cost in each configuration. Brackets in \textbf{b} and \textbf{e} show the total price--cost gap, $\Gamma$, the sourcing effect of competing demand, $\Delta c$, and the global-market price--cost difference, $\Delta^{pc}_{N}$. \textbf{c} shows production and}
    \label{fig:mechanism}
\end{figure}
\begin{figure}[t]
    \ContinuedFloat
    \caption*{allocation of Germany's suppliers in the two configurations. Band heights show each supplier's total hydrogen production, and ribbon widths show the volumes delivered to Germany and to other regions. Demand in other regions includes both consumption within the supplier's own region and deliveries to all other regions. \textbf{d--f} show the corresponding supply portfolios (\textbf{d,e}) and supplier allocation (\textbf{f}) for Japan. Colours identify suppliers across the diagrams. Germany and Japan have the same hydrogen demand in both configurations. Changes in the figures therefore reflect changes in sourcing and allocation rather than changes in delivered volume.}
\end{figure}

Germany's supply portfolio changes almost completely (Fig.~\ref{fig:mechanism}a--c). In the single-buyer configuration, Germany receives hydrogen from three external suppliers, with France alone providing 72\% of deliveries. In the global-market configuration, it receives hydrogen from six external suppliers. Deliveries from France fall from 228~TWh to zero, although French production increases from 228 to 350~TWh. Deliveries from Norway fall from 54 to 2~TWh, while its production also increases. The Baltic region, Poland, Iceland and Sweden, together with larger deliveries from Denmark, largely replace these suppliers, and the range of delivered costs widens from 109--129 to 116--188~EUR/MWh.

Japan's portfolio changes differently, as its main supplier stays the same, but provides a smaller share of Japan's supply (Fig.~\ref{fig:mechanism}d--f). CN-Northeast provides 98\% of Japan's hydrogen in the single-buyer configuration and 61\% in the global-market configuration. Its total production increases from 581 to 1,904~TWh, but only 19\% goes to Japan, while the rest serves domestic demand and other importers. Japan adds new suppliers, particularly Australia-East and New Zealand, and draws on more expensive supply blocks. The cost range widens from 122--150 to 120--192~EUR/MWh.

The two cases show a common reallocation mechanism, but in different forms. For Germany, activating demand elsewhere leads to an almost complete replacement of its main suppliers. Japan continues to rely on CN-Northeast, but receives a substantially smaller share of its output as that supplier serves demand in other regions. The following section examines how this changed portfolio contributes to the price--cost gap.

\subsection*{Most of the gap reflects price above average delivered cost}

\begin{figure}[!htbp]
    \centering
    \includegraphics[width=\textwidth]{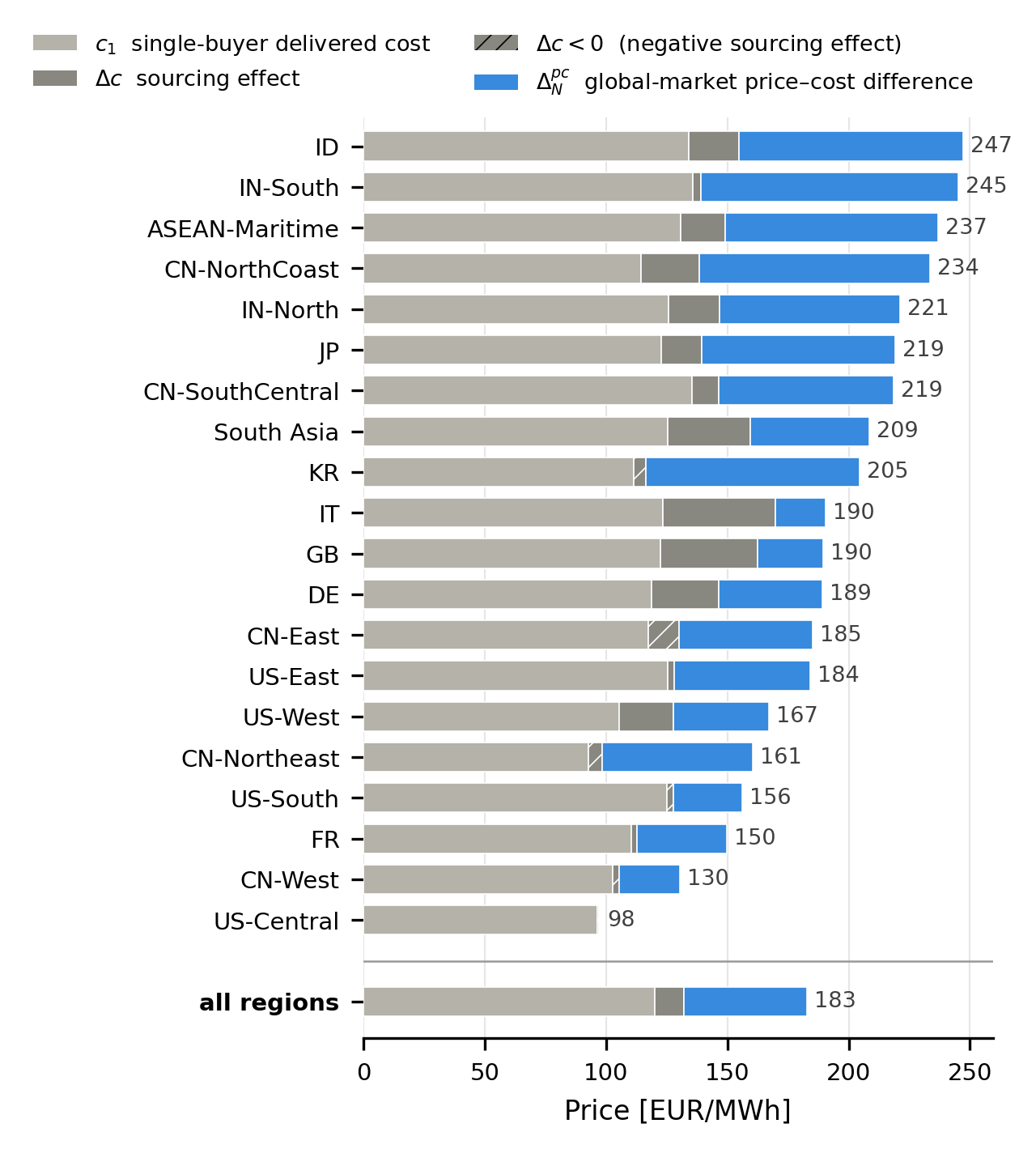}
    \caption{\textbf{Decomposition of global-market hydrogen prices in major demand regions in 2045.}
    Bars show the global-market price, $p_N$, for the 20 regions with the largest hydrogen demand in 2045, ordered by decreasing price, and with the demand-weighted result across the 81 analysed regions shown at the bottom. The price is decomposed into the single-buyer delivered-cost benchmark, $c_1$, the sourcing effect of competing demand, $\Delta c=c_N-c_1$, and the global-market price--cost difference, $\Delta^{pc}_{N}=p_N-c_N$. Hatching identifies regions in which $\Delta c<0$. Values at the ends of the bars report $p_N$ in EUR/MWh.}
    \label{fig:decomposition}
\end{figure}

Part of that price--cost gap reflects a more expensive delivered portfolio, but most of it does not (Fig.~\ref{fig:decomposition}). On a demand-weighted basis, the average delivered cost increases from 120 to 132~EUR/MWh between the two configurations, so the sourcing effect of competing demand, $\Delta c$, accounts for 12 of the 63~EUR/MWh gap. The remaining 51~EUR/MWh is the global-market price--cost difference, $\Delta^{pc}_{N}$. This difference arises because the price reflects the marginal value of additional supply, whereas the average delivered cost reflects the full supply portfolio. The two differ when an additional unit requires costlier production or transport, or when supplying one region means diverting hydrogen from another use, even under price-taking behaviour.

In 14 regions accounting for 30\% of global hydrogen demand, the sourcing effect $\Delta c$ is negative, indicating a lower global-market portfolio cost. In eight of these regions, accounting for 23\% of global demand, the wholesale price nevertheless rises. In the other six, both the average delivered cost and the wholesale price fall, most strongly in Canada, where the price drops by 42~EUR/MWh. Korea's average delivered cost falls from 116 to 111~EUR/MWh, while its price rises from 167 to 205~EUR/MWh. A cheaper delivered portfolio therefore does not necessarily imply a lower wholesale price. Germany and Japan have positive sourcing effects of 28 and 17~EUR/MWh and global-market price--cost differences of 43 and 80~EUR/MWh, respectively (Fig.~\ref{fig:decomposition}). Across the 78 regions with a positive gap, the relative sizes of the two components vary widely and do not appear systematically related to import status or net trade share.

\newpage

\section*{Discussion}

Comparing the two configurations reveals a large and persistent gap between single-buyer delivered-cost benchmarks and global-market prices. In 2045, the demand-weighted global-market price is 52\% above this benchmark. Most of this difference reflects prices above average delivered costs rather than a more expensive supply portfolio. The global-market price–cost difference exceeds the sourcing effect in 44 of the 78 positive-gap regions, accounting for 89\% of demand within that group.

Price--cost differences already occur in the single-buyer configuration. They do not require market power and can arise under price-taking behaviour when the marginal value of additional supply exceeds the average cost of the delivered portfolio. If all delivered volumes are valued at the regional marginal price, supply that costs less to produce and transport generates an inframarginal surplus. Under this interpretation, $\Delta^{pc}_{N}$ is the average surplus per delivered unit. The same mechanism can apply in self-sufficient regions. Once a region is connected to international trade, using low-cost hydrogen domestically means giving up the opportunity to sell it elsewhere, so domestic supply acquires an export opportunity value.

The most direct implication is for energy-system models that use exogenous regional hydrogen prices \cite{Schob.2023,Fleiter.2024,Moskalenko.2026}. Delivered-cost estimates are useful for comparing production locations and import routes, but they do not capture how market prices depend on the volume required, demand in other regions and the marginal value of supply. Using them as price assumptions can therefore understate the marginal price relevant to consumption and investment decisions, although published applications sometimes identify such cost-based paths explicitly as scenario or lower-bound assumptions rather than price forecasts \cite{Schmitz.2024}. Models that do not represent hydrogen-price formation endogenously should distinguish delivered-cost benchmarks from market-price assumptions and test region-specific price ranges across alternative demand, trade and infrastructure conditions. The demand-weighted gap found here should not be applied as a universal markup, because its size and composition vary substantially across regions (Figs.~\ref{fig:gap}~and~\ref{fig:decomposition}).

For importers, routes identified as low-cost in supply-chain studies may provide only part of the required volume once other buyers draw on the same production resources and transport options \cite{Hampp.2023}. Japan provides an illustrative policy case. Its strategy uses cost, insurance and freight (CIF) supply-cost targets, while also recognising international competition and pursuing partnerships and diversified supply chains \cite{MinisterialCouncilonRenewableEnergyofJapan.}. Qualitative recognition of competition and diversification does not by itself convert a supply-cost target into a market-price expectation. Japan's strategy also covers hydrogen derivatives, so its targets are not directly comparable with our modelled hydrogen prices. We therefore suggest that strategies should complement route-cost benchmarks with price scenarios that reflect the full required volume and competing demand. Long-term contracts, coordinated procurement and public infrastructure are not represented in our analysis, but may affect investment and price risk and should be assessed alongside these scenarios \cite{Antweiler.2024,Keutz.2025}.

For governments pursuing hydrogen exports, the results point to a potential tension between export promotion and access to low-cost hydrogen for domestic industry. Under the model's price-taking assumptions, opening access to foreign buyers can raise the opportunity value of domestic supply, so export growth and low-cost domestic decarbonisation should not be assumed to be automatically aligned. The United Arab Emirates illustrates this: as a large net exporter, its average delivered cost falls from 95 to 92~EUR/MWh between the two configurations while the wholesale price rises from 108 to 144~EUR/MWh. The model does not determine whether domestic production, imports or export restrictions are socially preferable, but it shows that these choices cannot be evaluated from delivered costs alone.

The modelled global-market prices range from 98 to 247~EUR/MWh among the major demand regions shown in Fig.~\ref{fig:decomposition}, illustrating that market integration does not necessarily eliminate regional price differences. Natural gas provides a useful comparison. In 2025, benchmark gas prices in Europe and Asia averaged around 12~USD/MMBtu, about 3.3 times the US benchmark of 3.6~USD/MMBtu \cite{IEA.2026.GasQ1}. Liquefaction and transport costs and infrastructure constraints can sustain such differences despite trade \cite{Oglend.2016,Farag.2025}. These regional price differences are distinct from the within-region price--cost gaps examined here.

The global-market configuration should be interpreted as a long-run reference rather than a forecast of realised hydrogen prices. It represents an integrated market with price-taking participants, whereas many announced low-emissions hydrogen projects rely on bilateral offtake arrangements \cite{InternationalEnergyAgency.2025}. Long-term contracting can generate price differences between markets and contract vintages \cite{Antweiler.2024}. The reported prices are marginal system values conditional on the assumed demand, investment pathway and resulting infrastructure, and should not be interpreted as short-term spot prices or expected contract prices. Whether the price--cost difference appears in actual payments, and who receives it, depends on market design and contracting arrangements. In non-convex markets, the payment rule can redistribute or eliminate the inframarginal surplus implied by the dual values \cite{Sioshansi.2014}, and under bilateral long-term contracts, delivered volumes do not need to be valued at a common marginal price at all \cite{Keutz.2025}. Such contracts can also affect investment and infrastructure use \cite{Keutz.2025}, but they do not necessarily eliminate the opportunity cost associated with constrained production and transport capacity \cite{Antweiler.2024}.

Market power is not included here, but can raise prices in importing regions relative to a competitive benchmark \cite{Barner.2024,Alanazi.2026}. In exporting regions, withholding volumes from export can leave lower-cost supply available domestically and reduce exporters' domestic prices \cite{Barner.2024}. How far sellers can raise prices also depends on buyers' alternatives. Pipeline infrastructure ties trading partners to particular routes, whereas shipping can offer greater flexibility. Market power should therefore not be treated as a uniform markup to the price--cost gap reported here.

Financing and contractual risk allocation can also change investment and prices. Lower financing costs can reduce equilibrium prices, although the magnitude depends on the resulting investment response \cite{Alanazi.2026,Antweiler.2024}. Contracts that shift risk to buyers, such as high take-or-pay rates, can impose costs on importing systems \cite{Keutz.2025}. Our region-specific financing assumptions use country risk premia and a common capital structure. They are neither technology-specific nor responsive to changing investor risk perceptions or de-risking policies. Financing assumptions and de-risking can materially alter whether a prospective exporter appears competitive \cite{Egli.2025,Liu.2026}.

The demand is exogenous, so the prices and gaps are conditional on the demand pathway. In sensitivity analyses with lower and higher demand (global 2045 demand 36\% below and 47\% above the base case), the demand-weighted global-market price is 4\% lower and 5\% higher, respectively, although regional changes are larger. The price--cost gaps assessed for Germany and Japan remain substantial (Supplementary Note~S5).
These tests do not replace price-responsive demand, under which quantities and prices adjust jointly \cite{Alanazi.2025,Alanazi.2026}.
Including additional carriers could alter delivered costs and trade routes, particularly for long-distance imports, where hydrogen derivatives are often a lower-cost transport option \cite{Moritz.2023,IRENA.2022b}. Their effect on the price--cost gap has not been tested here. The model also omits supplier-diversification, energy-security and geopolitical constraints. Japan's concentrated supply from CN-Northeast should therefore be interpreted as a least-cost allocation, not as a forecast of its procurement strategy.

In conclusion, delivered costs and market prices answer different questions and should not be treated as interchangeable. In the modelled global hydrogen market, the difference between them is large enough to matter for long-term price assumptions, investment decisions and infrastructure planning. Future work should test how price-responsive demand, additional carriers and strategic behaviour change these gaps, and how contracts and infrastructure ownership affect payments and the distribution of economic surplus. Access to low-cost hydrogen supply does not by itself guarantee a low regional market price.

\vspace{0.5cm}

\section*{Methods}

\setlength{\parindent}{0pt}

\subsection*{Hydrogen market framework}

We use the Hydrogen Price Model (HyPriM) to represent an interconnected global hydrogen market covering 81 regions between 2030 and 2050. Regions can produce, consume, import and export hydrogen, allowing market roles to emerge endogenously rather than being assigned a priori. Hydrogen is transported through pipelines and liquefied-hydrogen shipping infrastructure, both of which can be expanded endogenously by the model. The model determines production, trade, infrastructure investment and regional wholesale prices in a consistent optimisation framework.

HyPriM determines hydrogen prices using a two-stage optimization framework. In the first stage, the model minimizes total system cost through endogenous investment in hydrogen production and transport infrastructure under perfect foresight. Because some infrastructure choices are discrete, this stage is solved as a mixed-integer linear programme. In the second stage, installed production and transport capacities are fixed and the system is re-optimized as a linear programme using levelised production and transport cost coefficients. Regional wholesale prices are obtained from the dual variables of regional hydrogen balance constraints and represent the marginal value of supplying an additional unit of hydrogen at a given location. This use of duals from a linear pricing problem with fixed investment decisions is related to the approach of \citet{ONeill.2005} and to what \citet{Byers.2023} refer to as fixed-configuration pricing.

Under the model's price-taking assumptions, the cost-minimising allocation and associated balance duals from the fixed-capacity pricing stage provide a conditional competitive reference for the modelled hydrogen system. The dual values reflect marginal production and transport costs, binding production and infrastructure constraints, and the opportunity cost of allocating hydrogen to one region rather than another. Because capacities are fixed in the pricing stage, these dual values describe marginal prices for the installed system. They do not by themselves show whether revenues at those prices recover the costs of individual investments or act as forecasts of realized contract prices. A complete mathematical formulation of the investment stage, pricing stage and reported quantities is provided in the Supplementary Information.

\subsection*{Market configurations}

The analysis is designed as a matched comparison between two model configurations. In the \emph{global-market configuration}, all 81 regions have an active hydrogen demand and can simultaneously participate as hydrogen producers, consumers and traders. Prices therefore emerge from the system-wide allocation of hydrogen across finite production and transport opportunities.

For comparison, a separate \emph{single-buyer configuration} is solved for each region. In this configuration, demand is retained only for one focal region, while demand in all other regions is set to zero. The full set of global production potentials, infrastructure options and transport routes remains available. The focal region can therefore satisfy its prescribed demand through domestic production, imports, or both, but faces no competing demand from other regions. This setup is not a literal replication of an individual route study; rather, it extends route-level supply-chain logic to the focal region's full demand while removing competing demand. The single-buyer configuration is also not a monopsony: the focal region remains a price taker and exercises no buying power. Its wholesale price can therefore still exceed its average delivered cost when the marginal value of additional supply exceeds the average cost of the delivered portfolio.

All technology assumptions, renewable resource potentials, weighted average costs of capital, transport options and the focal region's demand pathway are identical between the matched single-buyer and global-market configurations. The configurations differ only in which demand regions are active. Differences in delivered portfolios and wholesale prices therefore arise from market-wide competing demand and the endogenous production, trade and infrastructure responses it induces.

The comparison is performed for all 81 modelled regions. Germany and Japan are used as detailed case studies because they are among the most frequently discussed future hydrogen importers and illustrate different forms of sourcing reallocation under competing demand. Germany is embedded in a densely interconnected continental market with access to both pipeline and maritime imports, whereas Japan can be connected by pipeline only to neighbouring mainland regions, so all other suppliers can only deliver to Japan by ship.

\subsection*{Representation of hydrogen supply}

Hydrogen production is represented through region-specific supply curves rather than single-point production costs. Each supply curve consists of multiple production blocks with increasing levelized costs of production and captures the finite availability of renewable energy resources within a region.

Supply curves are derived from weather-based assessments of renewable energy resources combined with an optimization of renewable generation and electrolysis capacity. Hydrogen production is assumed to rely on dedicated renewable resources. To avoid competition with direct electrification, projected domestic electricity demand is first allocated to the lowest-cost renewable resources, and only the remaining potential is available for hydrogen production.

For each supply-curve step, a separate optimization determines the cost-minimizing combination of renewable generation and electrolysis capacity required to produce a specified quantity of hydrogen. The resulting levelized costs therefore reflect both resource quality and endogenous technology sizing. Progressively larger hydrogen volumes can require the use of less favourable renewable resources, so supply costs generally rise with the volume produced.

The resulting supply curves differ substantially between regions. Resource-rich regions are characterized by long and relatively flat supply curves, whereas regions with limited renewable resources exhibit shorter and steeper curves. This heterogeneity allows marginal prices to differ from average delivered costs and shapes how competing demand affects regional prices and sourcing. Trade models that leave source capacity unconstrained cannot represent this heterogeneity, and the resulting trade patterns depend on that treatment \cite{Linsel.2023}.

A detailed description of the renewable resource assessment, supply-curve generation procedure and techno-economic assumptions is provided in Supplementary Note~S4.

\subsection*{Price metrics and decomposition}

To distinguish supply-chain cost benchmarks from market prices, we evaluate average delivered costs and wholesale prices for every focal region and year in both model configurations. The subscript \(1\) denotes the single-buyer configuration, in which only the focal region has active hydrogen demand. The subscript \(N\) denotes the global-market configuration, in which all \(N=81\) modelled regions have active hydrogen demand. Subscripts therefore refer to the number of active demand regions and not to region identifiers.

The average delivered cost in the single-buyer configuration, $c_1$, is the volume-weighted average production and transport cost of the hydrogen arriving at the focal region, including domestic production. Where hydrogen flows mix within the network, proportional allocation is used to associate upstream production and transport costs with the destination. The resulting portfolios are therefore accounting reconstructions rather than physical tracing of individual hydrogen volumes. We use $c_1$ as a system-level supply-chain cost benchmark. It retains the production-and-delivery cost perspective of route-level supply-chain analyses while extending it to the focal region's full demand: the model endogenously selects the cost-minimising combination of production blocks, suppliers, and transport routes while all other regions have zero
hydrogen demand. Consequently, $c_1$ is not intended to replicate the cost reported for any individual supply chain. The corresponding average delivered cost in the global-market configuration is denoted by $c_N$. The wholesale prices in the single-buyer and global-market configurations are denoted by $p_1$ and $p_N$, respectively, and are obtained from the dual value of the regional hydrogen balance in the corresponding pricing-stage optimisation.

The central outcome is the price--cost gap,
\[
\Gamma = p_N - c_1,
\]
which measures how far the global-market wholesale price exceeds the single-buyer supply-chain cost benchmark. The four quantities \(c_1\), \(p_1\), \(c_N\) and \(p_N\) form a cost-and-price table with two model configurations and two reported metrics. The same price--cost gap can therefore be reached along two paths through this table.

The first path moves first from average cost to price within the single-buyer configuration, and then from the single-buyer price to the global-market price:
\[
\Gamma = \underbrace{p_1 - c_1}_{\Delta^{pc}_{1}} + \underbrace{p_N - p_1}_{\Delta p}.
\]
Here, \(\Delta^{pc}_{1}\) is the single-buyer price--cost difference. It measures how far the single-buyer wholesale price exceeds the average delivered cost even when the focal region faces no competing demand. The term \(\Delta p\) is the price effect of competing demand. It measures how much the wholesale price changes when all other regions also demand hydrogen. This decomposition answers how much of the final price--cost gap is already present in the single-buyer benchmark and how the price changes when competing demand is introduced.

The second path moves first from the single-buyer average cost to the global-market average cost, and then from the global-market average cost to the global-market price:
\[
\Gamma = \underbrace{c_N - c_1}_{\Delta c} + \underbrace{p_N - c_N}_{\Delta^{pc}_{N}}.
\]
Here, \(\Delta c\) is the sourcing effect of competing demand. It measures how the average delivered cost changes when demand in other regions is activated, including changes in the supply portfolio and the timing of investment. The term \(\Delta^{pc}_{N}\) is the global-market price--cost difference. It measures how far the global-market wholesale price exceeds the average delivered cost of the portfolio that actually serves the focal region in the global-market configuration. This decomposition answers how much of the price--cost gap in the market outcome comes from a change in the cost of the delivered portfolio and how much comes from marginal pricing above the average delivered cost of that portfolio.

Both decompositions sum to the same price--cost gap, and we report them together because they answer different questions.
Because \(\Delta p\) alone is not the price--cost gap, \(\Gamma\) is reported alongside it throughout. Percentage values for \(\Gamma\) are expressed relative to \(c_1\). Absolute differences are used for regional distributions because relative measures can become unstable when benchmark values are small or when the difference changes sign. Differences and percentages are calculated from unrounded values.

International hydrogen trade is quantified by tracing hydrogen from its production region through the modelled transport network to the region of final consumption. At hubs receiving hydrogen from multiple origins, domestic production and imports are assumed to mix proportionally before being allocated to local consumption and onward transport. Hydrogen is counted as internationally traded when its production and consumption regions differ. Each unit is counted once at final consumption on an arrival basis, after transport and conversion losses, irrespective of how many borders or transit regions it crosses. The reporting threshold is 0.01~TWh.

\subsection*{Scenario assumptions and sensitivity analysis}

Hydrogen demand is specified exogenously and based on long-term global hydrogen deployment scenarios. Technology costs and transport parameters evolve over time according to the underlying scenario assumptions. All modelled costs and prices are expressed in constant 2025 euros. Hydrogen quantities  use the lower
heating value (LHV) basis. All regions can act as hydrogen producers, consumers, importers and exporters, with trade patterns emerging endogenously from the optimisation.

To test robustness, we repeat the global-market configuration under lower and higher demand scenarios and the single-buyer configuration for Germany and Japan. These analyses assess changes in regional wholesale prices and, for Germany and Japan, the price–cost gap and price effect of competing demand. Detailed scenario assumptions, techno-economic parameters and sensitivity specifications are provided in the Supplementary Information.

\subsection*{Data and code availability}

Input data and scenario definitions used in this study are documented in the
Supplementary Information. The resulting data will be made publicly available at Zenodo (DOI:~10.5281/zenodo.23780763) upon publication.

The code used to process the model results and reproduce the figures will be made available with publication.

\section*{Acknowledgements}

The authors thank Lukas Karkossa, Joshua Fragoso, Benjamin Lux, Katja Franke, Holger Lichtschläger and Robert Kunze for valuable discussions and data provision and Lukas Jansen for support with the WACC calculator tool \cite{Jansen_Global_WACC_Calculator_2026}. During the preparation of this work, the authors used DeepL to improve the clarity of the English writing as well as ChatGPT and Claude Opus to enhance the readability and language of the paper. After using these tools, the authors reviewed and edited the content as needed and take full responsibility for the content of the published article.

\section*{Funding}
B.W. and M.W. received funding for this work from the German Federal Ministry of Education and Research (BMBF) within the projects HYPAT (grant no. 03SF0620A) and HySecunda (grant no. 03SF0734), and from the Fraunhofer Cluster of Excellence Integrated Energy Systems (CINES). Open access funding provided by Fraunhofer-Gesellschaft zur Förderung der angewandten Forschung e.V.

\section*{Author Contributions}

Conceptualization: B.W., M.W., and R.M.; methodology: B.W.; software: B.W.; investigation/analysis: all authors; writing--original draft: B.W.; writing--review \& editing: all authors.

\section*{Competing interests}

The authors declare no competing interests.

\newpage

\bibliographystyle{unsrtnat}
\bibliography{references}

\newpage
\setcounter{section}{0}
\setcounter{figure}{0}
\setcounter{table}{0}
\setcounter{equation}{0}

\renewcommand{\thesection}{S\arabic{section}}
\renewcommand{\thefigure}{S\arabic{figure}}
\renewcommand{\thetable}{S\arabic{table}}
\renewcommand{\theequation}{S\arabic{equation}}

\titleformat{\section}[hang]
  {\normalfont\large\bfseries}
  {Supplementary Note \thesection\,\textbar}{0.45em}{}
\renewcommand{\thesubsection}{\arabic{section}.\arabic{subsection}}
\renewcommand{\thesubsubsection}{\thesubsection.\arabic{subsubsection}}
\renewcommand{\theequation}{S\arabic{equation}}
\captionsetup[figure]{name=Supplementary Figure, labelsep=period,
                      labelfont=bf, singlelinecheck=false, justification=justified}
\captionsetup[table]{name=Supplementary Table, labelsep=period,
                     labelfont=bf, singlelinecheck=false, justification=justified}
\setcounter{section}{0}\setcounter{figure}{0}\setcounter{table}{0}
\setcounter{equation}{0}

\section{Analysis setup and reported quantities}
\label{sec:si_setup}

This note defines the two model configurations and the costs, prices and decompositions reported in the main text. Supplementary Note~\ref{sec:si_hyprim} presents the mathematical formulation of the HyPriM market model.

\subsection{Model configurations}

The analysis compares two matched model configurations. Both use
identical technology costs, supply curves, weighted average costs of capital (WACCs), network topology and
transport options. At each active demand node, demand retains its set scenario value. The configurations differ only in which demand nodes are active. Production, trade, infrastructure investment and wholesale
prices are determined endogenously in both configurations.

\paragraph{Global-market configuration.}
In the global-market configuration, all 81 modelled regions have active
hydrogen demand. The demand parameter takes its scenario value at
every demand node,
\begin{equation}
q_{y,d} = q^{\mathrm{scen}}_{y,d}
\qquad \forall\, y \in Y,\ d \in D.
\label{eq:si_global_market}
\end{equation}
All regions can produce, consume and trade hydrogen simultaneously, drawing on the same global set of production potentials and transport options. The 81 modelled regions are listed in Supplementary Tables~\ref{tab:si_regions_countries} and~\ref{tab:si_regions_clusters}.

\paragraph{Single-buyer configuration for region \(i\).}
In the single-buyer configuration for region \(i\), demand is retained only
at the demand nodes belonging to region \(i\) and set to zero elsewhere,
\begin{equation}
q_{y,d} =
\begin{cases}
q^{\mathrm{scen}}_{y,d}, & h(d) = i,\\[0.3em]
0, & h(d) \neq i,
\end{cases}
\qquad \forall\, y \in Y,\ d \in D,
\label{eq:si_singlebuyer}
\end{equation}
where \(h(d)\) denotes the region to which demand node (d) belongs. The full global set of supply nodes, arcs and production potentials remains
available. The model therefore builds and operates the cost-minimal
production and delivery system serving region \(i\) alone, without demand from other regions. The benchmark extends route-level supply-chain analysis to the combination of suppliers and transport routes needed to meet the region’s full demand.

\subsection{Reported costs, prices and decompositions}

For each region \(i\) and year \(y\), we derive average delivered costs and
wholesale prices from the two matched configurations. Let \(c_{1,y,i}\) and
\(p_{1,y,i}\) denote the average delivered cost and wholesale price in the
single-buyer configuration for region \(i\). Let \(c_{N,y,i}\) and
\(p_{N,y,i}\) denote the corresponding quantities in the global-market
configuration, where \(N=81\) is the number of active demand regions. The
subscripts \(1\) and \(N\) therefore refer to the number of active demand
regions, not to region identifiers.

The modelled wholesale prices are given by the dual values of the regional hydrogen-balance constraints in Stage~2,
\begin{equation}
p_{1,y,i} = \lambda^{\mathrm{single}}_{y,i},
\qquad
p_{N,y,i} = \lambda^{\mathrm{global}}_{y,i}.
\label{eq:si_prices_reported}
\end{equation}
The average delivered costs \(c_{1,y,i}\) and \(c_{N,y,i}\) are calculated
in post-processing as the volume-weighted averages of the production and transport costs associated with all hydrogen delivered to region \(i\) in the respective
configuration, including domestic production.

The central reported quantity is the price--cost gap,
\begin{equation}
\Gamma_{y,i} = p_{N,y,i} - c_{1,y,i},
\label{eq:si_gap}
\end{equation}
which is the difference between the global-market wholesale price and the single-buyer delivered-cost benchmark. The four quantities
\(c_{1,y,i}\), \(p_{1,y,i}\), \(c_{N,y,i}\) and \(p_{N,y,i}\) form a
$(2\times2)$ table spanning two configurations and two reported metrics. The same
price--cost gap can therefore be decomposed along two paths through this
table.

The first path moves from average delivered cost to wholesale price within the
single-buyer configuration, and then from the single-buyer price to the
global-market price:
\begin{equation}
\Gamma_{y,i}
= \underbrace{p_{1,y,i} - c_{1,y,i}}_{\Delta^{pc}_{1,y,i}}
+ \underbrace{p_{N,y,i} - p_{1,y,i}}_{\Delta p_{y,i}} .
\label{eq:si_decomp_benchmark}
\end{equation}
Here, $\Delta^{pc}_{1,y,i}$ is the single-buyer price--cost difference, measuring the difference between the wholesale price and average delivered cost when only region \(i\) has active demand. The term \(\Delta p_{y,i}\) is the
price effect of competing demand, measuring how much the wholesale price changes when
demand in the other regions is activated. This decomposition shows how much
of the final price--cost gap is already present in the single-buyer
benchmark and how much is attributable to the price effect of competing demand.

The second path moves from the single-buyer average delivered cost to the
global-market average cost, and then from the global-market delivered
average cost to the global-market price:
\begin{equation}
\Gamma_{y,i} = \underbrace{c_{N,y,i} - c_{1,y,i}}_{\Delta c_{y,i}}
             + \underbrace{p_{N,y,i} - c_{N,y,i}}_{\Delta^{pc}_{N,y,i}} .
\label{eq:si_decomp_market}
\end{equation}
Here, \(\Delta c_{y,i}\) is the sourcing effect of competing demand, measuring how the
average delivered cost changes when demand in the other regions is activated. The term $(\Delta^{pc}_{N,y,i})$ is the global-market price--cost difference, measuring the difference between the global-market wholesale price and the average delivered cost of the portfolio that
 serves region \(i\) in the global-market configuration. This decomposition
shows how much of the overall price--cost gap is associated with a change in the delivered portfolio's average cost and how much with the difference between the global-market price and that average cost.

Both decompositions sum to the same price--cost gap, but separate it into different components to answer different questions. Neither decomposition is treated as primary. The price effect of competing demand, $\Delta p_{y,i}$, is always reported together with $\Gamma_{y,i}$, since it captures the change in price when demand in the other regions is activated, rather than the full price--cost gap.

The price--cost gap is also reported relative to the single-buyer average delivered-cost benchmark,
\begin{equation}
\gamma_{y,i} = \frac{\Gamma_{y,i}}{c_{1,y,i}} .
\label{eq:si_gap_relative}
\end{equation}
Absolute differences are used for regional distributions because relative measures can become large when benchmark values are small and can be harder to interpret when differences have different signs.

The two paths are linked by the identity
\begin{equation}
  \Delta p_{y,i} - \Delta c_{y,i} \;=\; \Delta^{pc}_{N,y,i} - \Delta^{pc}_{1,y,i},
  \label{eq:identity}
\end{equation}
which means that the difference between the price and sourcing effects equals the difference between the global-market and single-buyer price--cost differences.

The price effect of competing demand is expressed relative to the single-buyer
price,
\begin{equation}
\pi_{y,i} = \frac{\Delta p_{y,i}}{p_{1,y,i}}.
\label{eq:si_price_effect_relative}
\end{equation}

\section{Scenarios and model inputs}

\subsection{Demand data and scenarios}
Hydrogen and electricity demand are exogenous inputs to HyPriM. The underlying country-level demand trajectories are taken from the StEAM dataset \cite{Linsel.2025}, which is documented in the StEAM final project report \cite{Linsel.2025b}. In the StEAM dataset, country-level hydrogen and electricity demands are derived from the IEA World Energy Outlook \cite{InternationalEnergyAgency.2022}. Where the World Energy Outlook reports aggregated regions rather than individual countries, the StEAM dataset supplements the country-level estimates using additional literature and fills remaining gaps according to countries' shares of gross domestic product. For countries represented by multiple subnational regions in HyPriM, we distribute the national StEAM demand trajectories among the corresponding model regions. Apart from this spatial disaggregation, yearly values are interpolated linearly between the dataset’s reporting years, and demand is ramped up linearly in the early years reaching 2033.

The main results are based on the Renewable Electrification Scenario (RES), in which hydrogen is deployed in sectors where direct electrification is not feasible. The two alternative demand pathways from the same dataset are used for sensitivity analysis. The Announced Pledges Scenario (APS) assumes full implementation of existing climate commitments, with hydrogen consumption increasing later and remaining lower than in RES. Global hydrogen demand in 2045 is 36\% below RES, although the difference varies across regions. The Hydrogen Run-Up Scenario (HRU) represents stronger global hydrogen uptake, with high demand across power, industry, transport and heating. Global hydrogen demand in 2045 is 47\% above RES.

\subsection{Techno-economic and transport inputs}

The single-buyer and global-market configurations use the same
techno-economic assumptions and renewable resource potentials. Renewable resource potentials, full-load hours, hourly generation profiles and renewable-generation cost assumptions are taken from \cite{Franke.2024}. These inputs are used to construct the regional hydrogen supply curves described in Supplementary Note~\ref{sec:si_supply}. The principal technology assumptions and their
sources are summarised in Supplementary Table~\ref{tab:si_tech_inputs}.

For pipeline connections, we use route-specific distances and geographic
cost factors based on \cite{Weienburger.2026}. Pipeline investment costs
depend on the selected pipeline size, its cost per kilometre, the routed
distance and the route-specific cost factor. The model determines which
connections and capacities to build; it does not optimise the geographic
path of a pipeline.

Maritime route distances between the modelled ports were calculated
using the \texttt{searoute} Python package \cite{Halili.2026}.
The distances enter the calculation of round-trip times, which,
together with vessel capacity, determine how much shipping capacity
is needed for a given flow. Vessel and terminal costs, shipping losses
and other technical assumptions are specified separately.

\subsection{Financing assumptions}

Country-level real weighted average costs of capital (WACCs) were obtained
with the Global WACC Calculator \cite{Jansen_Global_WACC_Calculator_2026}. Supplementary
Table~\ref{tab:si_wacc_method} lists the settings used to calculate these WACCs. Subnational areas use the WACC of their country. For model
regions formed from multiple areas, we use the unweighted mean of the
constituent areas' rates. The values applied to all 81 model regions are
reported in Supplementary Table~\ref{tab:si_wacc_regions}.

\section{Mathematical formulation of the HyPriM market model}
\label{sec:si_hyprim}

\subsection{Overview and rationale}

HyPriM determines investment, operation and wholesale hydrogen prices in two sequential optimisation
stages.

\paragraph{Stage 1 (investment).}
A cost-minimising capacity expansion problem determines endogenous
investment in electrolysis, pipelines, liquefaction and regasification
terminals, and shipping capacity over the period 2030--2050 under perfect
foresight, subject to the demand balance in every region and year, with unmet demand penalised (Section~3.3.2).
Because pipelines are represented as discrete investments in three diameter classes, this stage is a mixed-integer linear program (MILP).

\paragraph{Stage 2 (pricing).}
All capacities are fixed at their Stage~1 values and system operation is re-optimised using the long-run marginal cost (LRMC) of each technology. With capacities fixed, the resulting problem is a linear program
(LP). The dual value of the hydrogen balance constraint at hub \(h\) in
year \(y\), denoted \(\lambda_{y,h}\), is reported as the modelled regional wholesale
price.

The separate pricing stage is needed because discrete investment decisions make the Stage~1 value function non-convex, meaning that its marginal values do not generally have the standard interpretation of market prices. Stage~2 fixes these investment decisions and yields a convex linear program for which LP duality holds. The hydrogen-balance dual values can therefore be interpreted as marginal system values. Conditional on the fixed Stage~1 capacities, the price vector \(\{\lambda_{y,h}\}\) together with the
optimal primal allocation constitutes a competitive equilibrium of the
modelled hydrogen market.

All monetary values are expressed in 2025 euros.
Hydrogen quantities are expressed in TWh per year on a lower heating value (LHV) basis. \footnote{HyPriM is formulated for a general set of commodities and can also represent synthetic aviation fuel and its associated conversion step. This study models only hydrogen, and demand for all other commodities is set to zero.
The commodity index is therefore suppressed throughout this section.}

\subsection{Notation and network structure}
\label{sec:si_notation}

\begin{figure}[!htbp]
\centering
\includegraphics[width=\textwidth]{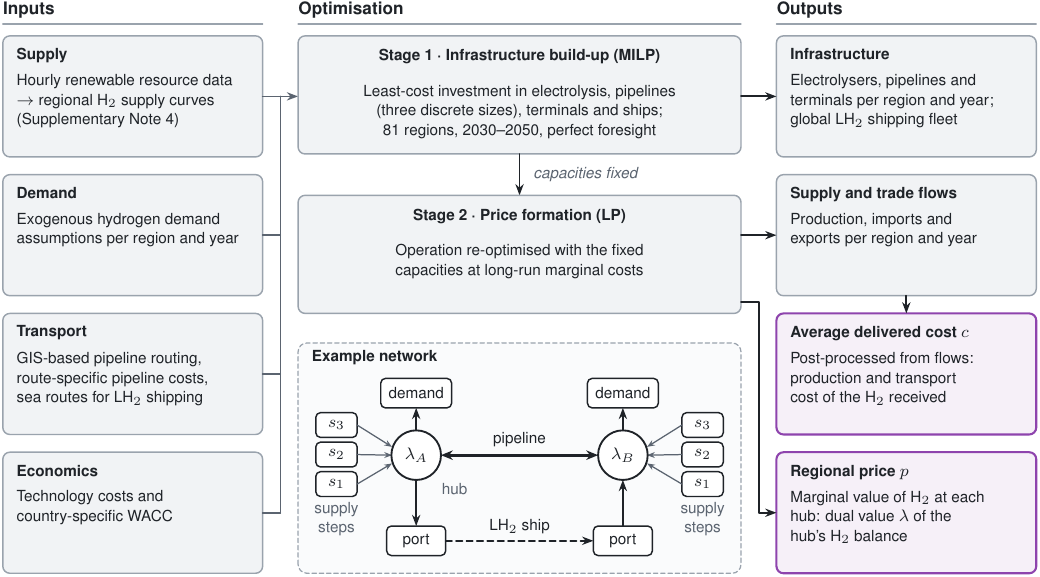}
\caption{\textbf{Structure of the HyPriM hydrogen-market model.}
 Regional hydrogen supply curves, exogenous demand, transport routes and economic assumptions enter a two-stage optimisation covering 81 regions from 2030 to 2050. Stage 1 determines least-cost investment in hydrogen production, pipelines, liquefaction and regasification terminals, and a global fleet of liquefied-hydrogen ships under perfect foresight. Pipelines are available in three discrete sizes, making this stage a mixed-integer linear programme (MILP). The resulting infrastructure capacities form the first output and are passed as fixed inputs to Stage 2. This linear programme (LP) re-optimises operation using long-run marginal cost coefficients and determines regional production, import and export flows. The average delivered cost, $c$, is calculated from these flows and the associated production and transport costs, and the regional wholesale price, $p$, is the dual value $\lambda$ of the hub’s hydrogen balance. These two quantities are highlighted in purple because they are the focus of this study. The example network shows two regional hubs, each connected to production supply steps, a demand node and a port. Hydrogen can move between hubs through pipelines or between ports by ship including conversion. $\lambda_A$ and $\lambda_B$ denote the regional hydrogen-balance dual values. The mathematical formulation is provided in Supplementary Note S3.}
\label{fig:si_market_model}
\end{figure}

The model sets and indices are listed in Supplementary
Table~\ref{tab:si_sets}, the transport arc sets in
Supplementary Table~\ref{tab:si_arcs}, the decision variables in
Supplementary Table~\ref{tab:si_vars} and the parameters in Supplementary
Table~\ref{tab:si_params}. Supplementary Figure~\ref{fig:si_market_model}
gives a schematic overview of the model chain, the network structure and
the two configurations used in the analysis. The network is built from six arc sets. For each
arc \(a\), \(o(a)\) denotes its origin and \(e(a)\) its destination
node.

\noindent Each pair of hubs that can be connected by pipeline is represented by exactly one pipeline
arc with an arbitrary orientation. The flow on the arc can take either sign: a
negative value of \(f_{y,a}\) denotes transport from \(e(a)\) to \(o(a)\). A
single physical corridor may host several parallel pipelines, which are
counted by the integer variables \(n_{y,a,z}\) introduced below.

\noindent All investment variables denote additions in year \(y\).
Available capacity is the sum of additions that remain within their
technical lifetime. Pipelines are added in discrete size classes, so
the same rule is applied to the number of installed pipes:
\begin{equation}
\hat{K}^{\theta}_{y,\cdot}
=
\sum_{v=\max(y_0,\,y-L^{\theta}_y+1)}^{y}
K^{\theta}_{v,\cdot},
\qquad
\hat{n}_{y,a,z}
=
\sum_{v=\max(y_0,\,y-L^{\mathrm{pipe}}_y+1)}^{y}
n_{v,a,z},
\qquad
\theta \in \{\mathrm{ely},\mathrm{liq},\mathrm{reg},\mathrm{shp}\}.
\label{eq:si_vintage}
\end{equation}
The pipeline capacity available on a route is obtained by multiplying
the installed number in each size class by its throughput capacity
(Eq.~\ref{eq:si_pipe_cap}). Within the 2030--2050 modelling horizon,
retirement affects ships, which have a lifetime of 15 years, but not
the other technologies.

\noindent The region-specific WACCs \(\rho_h\) are listed in
Supplementary Table~\ref{tab:si_wacc_regions}. For pipeline arcs, the
applicable WACC is the distance-weighted average of the country-specific
rates along the route. Shipping uses a uniform global WACC of 7\%,
reflecting the internationally mobile nature of shipping assets.

\paragraph{Annual costs over the remaining horizon.}
The annualised investment cost
\(\alpha^{\theta}_y c^{\theta,\mathrm{inv}}\) and fixed cost
\(c^{\theta,\mathrm{fix}}\) are multiplied by \(\omega^{\theta}_y\), the
number of years in which the asset serves the system within the horizon,
capped at its technical lifetime. An asset commissioned in 2030 is thus
charged for up to 21 years, depending on its technical lifetime, whereas
one commissioned in 2049 is charged for only two. Without this factor,
the model would favour early investment because capacity built near the
end of the horizon would lead to the same total cost while serving the
system for fewer years. The lifetime cap applies to ships commissioned
early in the horizon, it does not affect the other technologies within
the modelled period.

\subsection{Stage 1: infrastructure model}

\subsubsection{Objective}

Stage~1 minimises total system cost over the model horizon,
\begin{equation}
\min \quad \mathrm{TC} = \sum_{y \in Y} \big( C^{\mathrm{inv}}_{y} + C^{\mathrm{op}}_{y} \big),
\label{eq:si_obj_infra}
\end{equation}
The capacity-related cost assigned to investments made in year \(y\) is
\begin{align}
C^{\mathrm{inv}}_{y} =
&\sum_{s \in S} K^{\mathrm{ely}}_{y,s}\,
   \Big( \alpha^{\mathrm{ely}}_{y}\, c^{\mathrm{ely,inv}}_{y,s} + c^{\mathrm{ely,fix}}_{y,s} \Big)\, \omega^{\mathrm{ely}}_{y}
\nonumber \\[0.2em]
&+ \sum_{a \in \mathcal{A}^{\mathrm{pipe}}} \sum_{z \in Z} n_{y,a,z}\,
   \Big( \alpha^{\mathrm{pipe}}_{y}\, c^{\mathrm{pipe,inv}}_{y,a,z} + c^{\mathrm{pipe,fix}}_{y,a,z} \Big)\, \omega^{\mathrm{pipe}}_{y}
\nonumber \\[0.2em]
&+ \sum_{p \in P} \Big[ K^{\mathrm{liq}}_{y,p}
   \Big( \alpha^{\mathrm{liq}}_{y}\, c^{\mathrm{liq,inv}}_{y,p} + c^{\mathrm{liq,fix}}_{y,p} \Big)\, \omega^{\mathrm{liq}}_{y}
   + K^{\mathrm{reg}}_{y,p}
   \Big( \alpha^{\mathrm{reg}}_{y}\, c^{\mathrm{reg,inv}}_{y,p} + c^{\mathrm{reg,fix}}_{y,p} \Big)\, \omega^{\mathrm{reg}}_{y} \Big]
\nonumber \\[0.2em]
&+ K^{\mathrm{shp}}_{y}\,
   \Big( \alpha^{\mathrm{shp}}_{y}\, c^{\mathrm{shp,inv}}_{y} + c^{\mathrm{shp,fix}}_{y} \Big)\, \omega^{\mathrm{shp}}_{y},
\label{eq:si_capex}
\end{align}
The operating cost incurred in year \(y\) is
\begin{align}
C^{\mathrm{op}}_{y} =
&\sum_{p \in P} l_{y,p}\, e^{\mathrm{liq}}_{y}\, c^{\mathrm{el}}_{y}
+ \sum_{p \in P} r_{y,p}\, e^{\mathrm{reg}}_{y}\, c^{\mathrm{el}}_{y}
+ \sum_{a \in \mathcal{A}^{\mathrm{pipe}}} \tilde{f}_{y,a}\, c^{\mathrm{pipe,var}}_{y,a}
\nonumber \\[0.2em]
&+ \sum_{a \in \mathcal{A}^{\mathrm{ship}}} f_{y,a}\, c^{\mathrm{ship,var}}_{y,a}
+ \sum_{d \in D} u_{y,d}\, \varphi_{y,d}.
\label{eq:si_opex}
\end{align}

Electrolysis and its dedicated renewable generation are capital-intensive, with negligible short-run marginal costs. The model therefore  includes capital expenditure (CAPEX) \(c^{\mathrm{ely,inv}}\) and annual fixed O\&M costs \(c^{\mathrm{ely,fix}}\), with no separate variable production cost. Both are expressed per unit of annual hydrogen-production capacity (TWh/a) and are derived from the regional supply curves described in Supplementary Note~\ref{sec:si_supply}. The remaining variable costs arise from transport and conditioning. They comprise compression electricity for pipeline
transport,
\(c^{\mathrm{pipe,var}}_{y,a}
= \delta_a\,e^{\mathrm{pipe}}_{y}\,c^{\mathrm{el}}_{y}\),
electricity for liquefaction,
\(e^{\mathrm{liq}}_{y}c^{\mathrm{el}}_{y}\),
electricity for regasification,
\(e^{\mathrm{reg}}_{y}c^{\mathrm{el}}_{y}\),
and voyage costs for shipping.

\subsubsection{Node balances}

At each demand node, hydrogen deliveries and the unmet-demand slack variable \(u_{y,d}\) must equal exogenous demand scaled by the adequacy margin \(\mu\)(Section~\ref{sec:si_adequacy}). The slack variable is penalised at \(\varphi_{y,d}\):
\begin{equation}
\sum_{a \in \mathcal{A}^{\mathrm{hd}}:\, e(a)=d} f_{y,a} + u_{y,d}
= \mu\, q_{y,d}
\qquad \forall\, y \in Y,\ d \in D.
\label{eq:si_dem_infra}
\end{equation}

The hydrogen balance at each hub accounts for production, bidirectional pipeline flows, regasified imports, hydrogen sent to liquefaction, and
deliveries to demand:
\begin{align}
&\sum_{a \in \mathcal{A}^{\mathrm{sh}}:\, e(a)=h} f_{y,a}
+ \sum_{a \in \mathcal{A}^{\mathrm{ph}}:\, e(a)=h} f_{y,a}
+ \sum_{a \in \mathcal{A}^{\mathrm{pipe}}:\, e(a)=h} f_{y,a}
\nonumber \\
&\qquad
- \sum_{a \in \mathcal{A}^{\mathrm{pipe}}:\, o(a)=h} f_{y,a}
- \sum_{a \in \mathcal{A}^{\mathrm{hp}}:\, o(a)=h} f_{y,a}
- \sum_{a \in \mathcal{A}^{\mathrm{hd}}:\, o(a)=h} f_{y,a}
= 0
\qquad \forall\, y \in Y,\ h \in H.
\label{eq:si_hub_bal}
\end{align}
Pipeline flows are assumed to incur no physical hydrogen losses. Compression electricity is instead included in the variable transport cost \(c^{\mathrm{pipe,var}}_{y,a}\).

Production, liquefaction and regasification quantities are linked to flows on the corresponding arcs:
\begin{align}
f_{y,a} &= g_{y,s} && \forall\, y,\ a=(s,h) \in \mathcal{A}^{\mathrm{sh}}, \label{eq:si_sup_bal}\\
f_{y,a} &= l_{y,p} && \forall\, y,\ a=(h,p) \in \mathcal{A}^{\mathrm{hp}}, \label{eq:si_liq_bal}\\
f_{y,a} &= r_{y,p} && \forall\, y,\ a=(p,h) \in \mathcal{A}^{\mathrm{ph}}. \label{eq:si_reg_bal}
\end{align}

At each port, total outgoing shipping flow equals the volume liquefied, while total incoming flow after transport losses equals the volume regasified:
\begin{align}
\sum_{a \in \mathcal{A}^{\mathrm{ship}}:\, o(a)=p} f_{y,a} &= l_{y,p}
&& \forall\, y \in Y,\ p \in P, \label{eq:si_port_out}\\
\sum_{a \in \mathcal{A}^{\mathrm{ship}}:\, e(a)=p} \big(1-\ell_{y,a}\big) f_{y,a} &= r_{y,p}
&& \forall\, y \in Y,\ p \in P. \label{eq:si_port_in}
\end{align}

\subsubsection{Adequacy margin}
\label{sec:si_adequacy}

In Stage~1, demand is scaled by an adequacy margin of 5\% \((\mu = 1.05)\)
(Eq.~\ref{eq:si_dem_infra}). Stage~2 clears at unscaled demand. This difference creates operational headroom in the fixed capacities passed to Stage~2. If Stage~1 instead sized the system exactly to demand, binding capacity constraints could make the resulting dual prices sensitive to the precise investment solution and, in limiting cases, to the slack penalty. The adequacy margin reduces this sensitivity, so prices are more often determined by marginal supply and transport options.

The margin is analogous to the planning reserve margin used in
electricity capacity-expansion models, but it is
applied to the annual energy balance rather than to a separate peak-capacity
adequacy constraint, which is done here because the model has annual
resolution and no intra-year peak. Because \(\mu\) determines the available headroom, it also affects the shadow values of binding capacity constraints.

\subsubsection{Capacity constraints}

Hydrogen output from each supply-curve step is limited by installed electrolysis capacity, which cannot exceed the available potential of that step:
\begin{align}
g_{y,s} &\le \hat{K}^{\mathrm{ely}}_{y,s} && \forall\, y \in Y,\ s \in S, \label{eq:si_gen_cap}\\
\hat{K}^{\mathrm{ely}}_{y,s} &\le \bar{G}_{y,s} && \forall\, y \in Y,\ s \in S. \label{eq:si_pot}
\end{align}

Annual hydrogen production at each hub can increase by at most $\bar{\Delta}_{y,h}$ from one year to the next:
\begin{align}
\sum_{s \in S(h)} g_{y,s}
&\le \sum_{s \in S(h)} g_{y-1,s} + \bar{\Delta}_{y,h}
&& \forall\, y \in Y \setminus \{y_0\},\ h \in H.
\label{eq:si_rampcap}
\end{align}

Liquefaction and regasification throughput at each port cannot exceed the corresponding installed capacity:
\begin{align}
l_{y,p} &\le \hat{K}^{\mathrm{liq}}_{y,p} && \forall\, y \in Y,\ p \in P, \label{eq:si_liq_cap}\\
r_{y,p} &\le \hat{K}^{\mathrm{reg}}_{y,p} && \forall\, y \in Y,\ p \in P. \label{eq:si_reg_cap}
\end{align}

Bidirectional pipeline flow \(f_{y,a}\) may take either sign. The auxiliary
variable \(\tilde{f}_{y,a}\) bounds its absolute magnitude and is used to
calculate variable transport costs and enforce the capacity limit:
\begin{align}
f_{y,a} \le \tilde{f}_{y,a}, \qquad
-f_{y,a} &\le \tilde{f}_{y,a}
&& \forall\, y \in Y,\ a \in \mathcal{A}^{\mathrm{pipe}},
\label{eq:si_abs}\\
\tilde{f}_{y,a}
&\le \sum_{z \in Z} \hat{n}_{y,a,z}\,
\kappa^{\mathrm{pipe}}_{y,z}
&& \forall\, y \in Y,\ a \in \mathcal{A}^{\mathrm{pipe}}.
\label{eq:si_pipe_cap}
\end{align}

At most \(\bar{n}\) pipelines may be added to each arc in each year, representing the construction-rate limit for a single corridor:
\begin{equation}
\sum_{z \in Z} n_{y,a,z} \le \bar{n}
\qquad \forall\, y \in Y,\ a \in \mathcal{A}^{\mathrm{pipe}}.
\label{eq:si_pipe_qty}
\end{equation}

Shipping is represented by a single global fleet. Flow on each route occupies cargo capacity for its round-trip time, and total use of shipping cannot exceed the fleet's annual availability:
\begin{equation}
\sum_{a \in \mathcal{A}^{\mathrm{ship}}}
f_{y,a}\, \tau_{y,a}
\le
8760 \cdot \hat{K}^{\mathrm{shp}}_{y}
\qquad \forall\, y \in Y.
\label{eq:si_ship_cap}
\end{equation}
Longer routes therefore require more fleet capacity for the same annual transport volume, without assigning individual vessels to specific routes.

\subsection{Stage 2: price model}

\subsubsection{Formulation}

All capacity variables determined in Stage~1 are fixed at their optimal
values \(\hat{K}^{\theta,\star}\) and \(\hat{n}^{\star}\) and enter
Stage~2 as parameters. Constraints
\eqref{eq:si_hub_bal}--\eqref{eq:si_port_in},
\eqref{eq:si_gen_cap},
\eqref{eq:si_liq_cap}--\eqref{eq:si_pipe_cap} and
\eqref{eq:si_ship_cap} are retained unchanged. The supply-potential constraint
\eqref{eq:si_pot} and pipeline construction-rate constraint
\eqref{eq:si_pipe_qty} are omitted because they restrict investment
decisions that are fixed in Stage~2. The production-growth constraint
\eqref{eq:si_rampcap} is also omitted because it applies only during
Stage~1 infrastructure expansion. The demand balance is imposed without
the adequacy margin:
\begin{equation}
\sum_{a \in \mathcal{A}^{\mathrm{hd}}:\, e(a)=d} f_{y,a} + u_{y,d}
= q_{y,d}
\qquad \forall\, y \in Y,\ d \in D.
\label{eq:si_dem_price}
\end{equation}

Stage~2 minimises total dispatch cost, with production, conditioning and transport charged at their respective long-run marginal costs and unmet demand charged at the penalty \(\varphi_{y,d}\):
\begin{align}
\min \quad \sum_{y \in Y} \Bigg[
&\sum_{s \in S} g_{y,s}\, \hat{c}^{\mathrm{ely}}_{y,s}
+ \sum_{p \in P} \Big( l_{y,p}\, \hat{c}^{\mathrm{liq}}_{y,p} + r_{y,p}\, \hat{c}^{\mathrm{reg}}_{y,p} \Big)
\nonumber \\
&+ \sum_{a \in \mathcal{A}^{\mathrm{pipe}}} \tilde{f}_{y,a}\, \hat{c}^{\mathrm{pipe}}_{y,a}
+ \sum_{a \in \mathcal{A}^{\mathrm{ship}}} f_{y,a}\, \hat{c}^{\mathrm{ship}}_{y,a}
+ \sum_{d \in D} u_{y,d}\, \varphi_{y,d}
\Bigg].
\label{eq:si_obj_price}
\end{align}

\paragraph{Pricing basis}

Investment and fixed O\&M account for a large share of the costs represented
in HyPriM. A pricing stage based only on short-run variable costs would
therefore omit much of the cost of hydrogen production and transport.
Stage~2 assigns each technology
a long-run marginal cost that includes annualised investment costs, fixed
O\&M and, where applicable, variable operating costs. The resulting dual
values are long-run marginal system values conditional on the capacities
selected in Stage~1.

\subsubsection{Long-run marginal cost coefficients}

\paragraph{Electrolysis}

Each electrolysis-capacity vintage at supply node \(s\) is assigned the levelised cost of hydrogen applicable in its commissioning year:
\begin{equation}
\mathrm{LCOH}_{v,s} = \alpha^{\mathrm{ely}}_{v}\, c^{\mathrm{ely,inv}}_{v,s} + c^{\mathrm{ely,fix}}_{v,s},
\label{eq:si_lcoh_vintage}
\end{equation}
This reflects the technology costs and regional WACC prevailing when the capacity is commissioned.

Stage~2 tracks hydrogen production by supply node and year, but not by construction vintage. It therefore assigns output at each node the capacity-weighted average LCOH of all vintages installed by year \(y\):
\begin{equation}
\hat{c}^{\mathrm{ely}}_{y,s}
=
\frac{\sum_{v \le y} K^{\mathrm{ely},\star}_{v,s}\, \mathrm{LCOH}_{v,s}}
     {\sum_{v \le y} K^{\mathrm{ely},\star}_{v,s}}
\qquad \text{if } \textstyle\sum_{v \le y} K^{\mathrm{ely},\star}_{v,s} > 0 .
\label{eq:si_lrmc_ely}
\end{equation}
This averaging does not remove differences among supply-curve steps. Each step \(s\) remains a separate node with its own LCOH, and Eq.~\eqref{eq:si_lrmc_ely} averages only across construction vintages within that step.

\paragraph{Terminals}

Liquefaction and regasification are charged at their respective annualised
investment, fixed O\&M and electricity costs:
\begin{align}
\hat{c}^{\mathrm{liq}}_{y,p}
&= \alpha^{\mathrm{liq}}_{y}\, c^{\mathrm{liq,inv}}_{y,p}
 + c^{\mathrm{liq,fix}}_{y,p}
 + e^{\mathrm{liq}}_{y}\, c^{\mathrm{el}}_{y},
\label{eq:si_lrmc_liq}\\
\hat{c}^{\mathrm{reg}}_{y,p}
&= \alpha^{\mathrm{reg}}_{y}\, c^{\mathrm{reg,inv}}_{y,p}
 + c^{\mathrm{reg,fix}}_{y,p}
 + e^{\mathrm{reg}}_{y}\, c^{\mathrm{el}}_{y}.
\label{eq:si_lrmc_reg}
\end{align}

\paragraph{Pipelines}

When pipelines of different sizes coexist on an arc, Stage~2 uses the pipeline-count-weighted average of their size-specific unit costs and adds compression electricity:
\begin{equation}
\hat{c}^{\mathrm{pipe}}_{y,a}
=
\frac{\displaystyle\sum_{z \in Z} \hat{n}^{\star}_{y,a,z}\,
      \frac{\alpha^{\mathrm{pipe}}_{y}\, c^{\mathrm{pipe,inv}}_{y,a,z} + c^{\mathrm{pipe,fix}}_{y,a,z}}
           {\kappa^{\mathrm{pipe}}_{y,z}}}
     {\displaystyle\sum_{z \in Z} \hat{n}^{\star}_{y,a,z}}
+ \delta_a\, e^{\mathrm{pipe}}_{y}\, c^{\mathrm{el}}_{y}.
\label{eq:si_lrmc_pipe}
\end{equation}
If no pipeline is installed on an arc, Eq.~\eqref{eq:si_pipe_cap} constrains its flow to zero.

\paragraph{Shipping.}

On route \(a\), the round-trip time \(\tau_{y,a}\) determines the number of
voyages a vessel can complete per year,
\(n^{\mathrm{trip}}_{y,a}=8760/\tau_{y,a}\), and its resulting annual
transport capacity,
\(E_{y,a}=\kappa^{\mathrm{shp}}_{y}n^{\mathrm{trip}}_{y,a}\).
The shipping cost per unit transported is therefore route-dependent:
\begin{equation}
\hat{c}^{\mathrm{ship}}_{y,a}
=
\frac{\alpha^{\mathrm{shp}}_{y}\,C^{\mathrm{shp,inv}}_{y}
      + C^{\mathrm{shp,fix}}_{y}
      + c^{\mathrm{trip}}_{y}\,n^{\mathrm{trip}}_{y,a}}
     {E_{y,a}},
\label{eq:si_lrmc_ship}
\end{equation}
where \(C^{\mathrm{shp,inv}}_{y}\) and
\(C^{\mathrm{shp,fix}}_{y}\) are the investment and annual fixed O\&M costs
per vessel, respectively, and \(c^{\mathrm{trip}}_{y}\) is the variable cost
per voyage.

\subsubsection{Cost components in Stage~2}

For all technologies, Stage~2 combines annualised investment costs, fixed O\&M costs and variable costs:
\begin{equation}
\hat{c}^{\theta} = \alpha^{\theta} c^{\theta,\mathrm{inv}} + c^{\theta,\mathrm{fix}} + c^{\theta,\mathrm{var}},
\label{eq:si_consistency}
\end{equation}
Where necessary, these costs are averaged across installed assets and
expressed per unit of annual output or transport, as described above. The
horizon factor \(\omega^{\theta}_{y}\) used in Stage~1 is omitted because
Stage~2 applies these annual unit costs separately in each model year after
capacities have been fixed.

\subsection{Wholesale price extraction}
\label{sec:si_prices}

In Stage~2, the investment decisions from Stage~1 are fixed, leaving a linear pricing problem. For each hub \(h\) and year \(y\), we use the dual variable \(\lambda_{y,h}\) of the hub balance \eqref{eq:si_hub_bal} as the regional wholesale price. It represents the marginal system value of one additional unit of hydrogen at that hub. By comparison, average delivered cost measures the volume-weighted production and transport cost of the hydrogen supplied to a region. Regional
price differences can therefore arise from transport costs, binding
production or transport capacities, and the opportunity cost of directing
supply to one region rather than another.

Unmet demand is represented by the slack variable \(u_{y,d}\), which is
charged at the penalty \(\varphi_{y,d}\). In the reported runs, the penalty
is set to \(\varphi_{y,d}=999~\mathrm{EUR/MWh}\). At hubs that supply an
attached demand node, this penalty provides an upper bound on the regional
price:
\begin{equation}
\lambda_{y,h}
\le
\max_{d:\,h(d)=h}\varphi_{y,d}.
\label{eq:si_slack_bound}
\end{equation}
If a demand node is supplied partly from the hub and partly through the
slack variable, the regional price equals the slack penalty. Observations
with penalty-determined prices are excluded from the price--cost
comparison. In the base case, unmet demand is most widespread in 2030–2031. No region has a penalty-determined price in 2045 in either configuration. Between 2035 and 2050, only Switzerland is affected, in 2041–2044 in the global-market configuration. In 2043–2044, it also sets the price in Austria, which supplies Switzerland through a pipeline with spare capacity, so that Austria’s price equals the penalty less the transport cost to Switzerland. These observations are excluded from the pathway statistics.

\subsection{Solution procedure}

The model is solved in two stages:
\begin{enumerate}
\item \textbf{Infrastructure stage.} The Stage~1 MILP is solved with a relative MIP gap tolerance of 1\%.
\item \textbf{Pricing stage.} The Stage~2 LP defined by Eqs.
\eqref{eq:si_obj_price} and \eqref{eq:si_dem_price} is solved with all
capacities fixed at their Stage~1 values. The dual variables of the hub
balances \eqref{eq:si_hub_bal} are then extracted.
\end{enumerate}
To reduce runtime, the Stage~1 MILP is warm-started from its LP relaxation,
in which the pipeline investment variables \(n_{y,a,z}\) are continuous.
This relaxation is a computational initialization rather than a separate
model stage. All optimisation problems are solved with Gurobi. Regional prices can vary across near-optimal capacity solutions, even when total system costs are similar.

\section{Hydrogen supply curve model}
\label{sec:si_supply}

\subsection{Purpose and scope}

The market model of Supplementary Note~\ref{sec:si_hyprim} represents each region's hydrogen production
potential as a set of discrete supply steps \(s \in S(h)\), each characterised
by an annual volume \(\bar{G}_{y,s}\) and specific costs
\(c^{\mathrm{ely,inv}}_{y,s}\) and \(c^{\mathrm{ely,fix}}_{y,s}\). This section
describes how these steps are derived.

For each region, we solve a separate optimisation for a series of increasing annual hydrogen production targets. The model selects the cost-minimal combination of renewable generation and electrolysis capacity using hourly renewable-resource profiles. As lower-cost renewable potentials are exhausted, higher production targets require more expensive resource classes, producing an upward-sloping hydrogen supply curve. The curves are calculated for 2030 and 2050 and linearly interpolated for the intervening years.

\subsection{Notation}

Symbols in this section are local to it. In particular, \(h\) denotes hours
rather than hydrogen-market regions. All monetary values are expressed in
2025 EUR. Energy quantities are reported in MWh or MWh/a, capacities in MW,
investment costs in EUR/MW, fixed O\&M costs in EUR/(MW\,a), and levelised
costs in EUR/MWh.

\begin{figure}[!htbp]
\centering
\includegraphics[width=\textwidth]{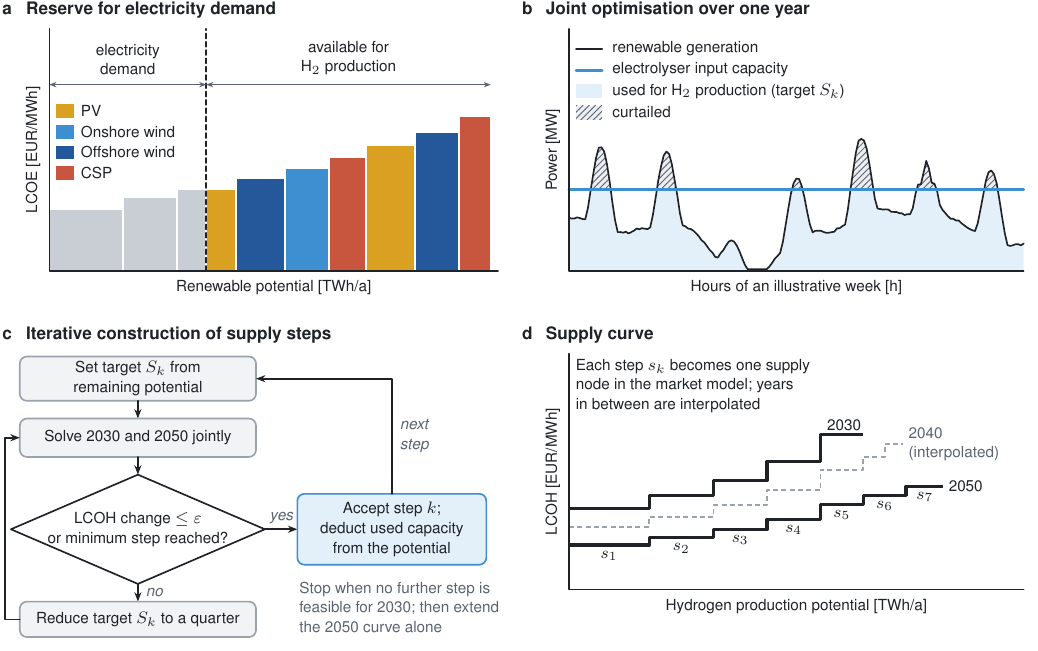}
\caption{\textbf{Construction of regional hydrogen supply curves.}
(a) Renewable resource classes are ordered by their levelised cost of electricity (LCOE). Grey bars show the potential reserved to meet regional electricity demand. The remaining potential is available for hydrogen production, with colours distinguishing the technology options solar photovoltaics (PV), onshore wind, offshore wind and concentrating solar power (CSP). (b) For an annual hydrogen production target $S_k$, an hourly optimisation determines renewable generation and electrolyser capacities jointly. The example week shows how the optimised mix of renewable technologies supplies electricity to meet the annual hydrogen production target, with electricity use limited by the optimised electrolyser input capacity and excess generation curtailed. The optimisation covers a full year. (c) shows how the hydrogen supply curve is built step by step by repeating the optimisation shown in b for successive hydrogen production targets. The 2030 and 2050 systems are optimised jointly, with the same renewable and electrolyser capacities in both years but year-specific costs. A step is accepted when the change in its 2030 levelised cost of hydrogen (LCOH) relative to the preceding step is within the threshold $\varepsilon$, or when the minimum step size is reached. Otherwise, the production target is reduced to one quarter and the optimisation is repeated. After each accepted step, the renewable capacity used is deducted from the remaining potential. Once no further step is feasible for 2030, the procedure continues for 2050 alone, where additional renewable potential, for example due to increased efficiency, could allow further supply steps. (d) The resulting supply curves show the hydrogen production potential available at different costs. Each step represents an additional potential supply volume and its associated production cost. Costs and available volumes are interpolated between 2030 and 2050, illustrated here for 2040. Each step represents a supply node step in the market model.}
\label{fig:si_supply_curves}
\end{figure}

The sets, indices and parameters of the supply-curve model are listed in
Supplementary Table~\ref{tab:si_s2_params}, and the decision variables in
Supplementary Table~\ref{tab:si_s2_vars}. Supplementary
Figure~\ref{fig:si_supply_curves} summarises the procedure. For each region, the regional cost of capital \(\rho\) is used to annualise the investment costs of both renewable generation and electrolysis, ensuring consistency with the market model.

\subsection{Reservation of renewable potential for domestic electricity demand}

Hydrogen production uses only the renewable potential remaining after
domestic electricity demand has been met. Before constructing the hydrogen
supply curve, Step~0 therefore reserves the cost-minimal renewable capacity
required to cover each region's electricity demand:
\begin{align}
\min_{K^{g,0}} \quad
& \sum_{y \in \mathcal{Y}}\sum_{t \in \mathcal{T}}
\big(
\alpha^{g}_{y,t}I^{g}_{y,t}
+ f^{g}_{y,t}
\big)K^{g,0}_{y,t}
\nonumber\\
\text{s.t.} \quad
& \sum_{t \in \mathcal{T}}K^{g,0}_{y,t}F_{y,t}
= D^{\mathrm{el}}
&& \forall\, y \in \mathcal{Y},
\nonumber\\
& 0 \le K^{g,0}_{y,t}
\le \bar{K}^{\mathrm{raw}}_{y,t}
&& \forall\, y \in \mathcal{Y},\ t \in \mathcal{T}.
\label{eq:si_s2_step0}
\end{align}

The renewable potential remaining for hydrogen production is
\begin{equation}
\bar{K}_{y,t}
=
\bar{K}^{\mathrm{raw}}_{y,t}
-
K^{g,0}_{y,t}
\qquad
\forall\, y \in \mathcal{Y},\ t \in \mathcal{T}.
\label{eq:si_s2_remaining}
\end{equation}

The regional electricity demand projected for 2050 in the underlying demand scenario is reserved in both supply-curve support years, 2030 and 2050. Hydrogen production is not coupled to the electricity grid and uses only the remaining dedicated renewable potential.

\subsection{Iterative supply-step construction}

For each supply-curve step \(k\), the model jointly determines renewable
generation and electrolysis capacities for a given annual hydrogen
production target \(S_k\), subject to the remaining renewable potentials
\(\bar{K}_{y,t}\):
\begin{equation}
\min \quad \sum_{y \in \mathcal{Y}} \left[
\sum_{t \in \mathcal{T}} \big( \alpha^{g}_{y,t} I^{g}_{y,t} + f^{g}_{y,t} \big) K^{g}_{y,t}
+ \big( \alpha^{e}_{y} I^{e}_{y} + f^{e}_{y} \big) K^{e}_{y}
\right]
\label{eq:si_s2_obj}
\end{equation}
subject to
\begin{align}
\sum_{t \in \mathcal{T}} g_{y,h,t} &= G_{y,h}
&& \forall\, y \in \mathcal{Y},\ h \in \mathcal{H}, \label{eq:si_s2_elbal}\\
\eta\, G_{y,h} &= Q_{y,h}
&& \forall\, y \in \mathcal{Y},\ h \in \mathcal{H}, \label{eq:si_s2_conv}\\
Q_{y,h} &\le K^{e}_{y}
&& \forall\, y \in \mathcal{Y},\ h \in \mathcal{H}, \label{eq:si_s2_elycap}\\
g_{y,h,t} &\le K^{g}_{y,t}\, \phi_{y,h,t}\, F_{y,t}
&& \forall\, y \in \mathcal{Y},\ h \in \mathcal{H},\ t \in \mathcal{T}, \label{eq:si_s2_avail}\\
K^{g}_{y,t} &\le \bar{K}_{y,t}
&& \forall\, y \in \mathcal{Y},\ t \in \mathcal{T}, \label{eq:si_s2_pot}\\
\sum_{h \in \mathcal{H}} Q_{y,h} &= S_k
&& \forall\, y \in \mathcal{Y}, \label{eq:si_s2_target}\\
K^{g}_{2030,t} &= K^{g}_{2050,t}
&& \forall\, t \in \mathcal{T},
\label{eq:si_s2_gen_coupling}\\
K^{e}_{2030} &= K^{e}_{2050}.
\label{eq:si_s2_ely_coupling}
\end{align}

Electrolysis capacity \(K^{e}_{y}\) is measured as the maximum hourly
hydrogen output. It therefore limits hydrogen production \(Q_{y,h}\),
rather than electricity input \(G_{y,h}\), in
Eq.~\eqref{eq:si_s2_elycap}.

Electricity that cannot be absorbed by the electrolyser is curtailed. The model jointly sizes renewable generation and electrolysis capacity, resulting in different electrolyser full-load hours across regions and supply-curve steps.

To ensure that each supply step represents the same physical system in both support years, 2030 and 2050 are optimised jointly. The coupling constraints require the same renewable capacity in each resource class and the same electrolysis capacity in both years. The composition of each step therefore remains unchanged, while its cost reflects the techno-economic assumptions for the respective year.

\subsection{Assembly of the regional supply curves}

Supply curve steps are generated sequentially. After each accepted step the renewable
capacity it uses is deducted from the remaining potential, so that the next
step must draw on the remaining resources. Step widths are not uniform,
they adapt to the size of the remaining resource classes and the
steepness of the curve.

For each region, the supply curve is assembled as follows:
\begin{enumerate}\setlength{\itemsep}{1pt}
\item \emph{Initialise.} Solve Eq.~\eqref{eq:si_s2_step0} and initialise the remaining renewable potentials using Eq.~\eqref{eq:si_s2_remaining}. Set \(k \leftarrow 0\).
\item \emph{Determine the production target.} For each resource class with remaining potential, calculate its annual electricity generation \[ E_t = \bar{K}_{2030,t}F_{2030,t} \] and its levelised electricity cost \[ \mathrm{LCOE}_t = \frac{ \alpha^{g}_{2030,t}I^{g}_{2030,t} + f^{g}_{2030,t} }{ F_{2030,t} }. \] Let \(E_{(1)}\) and \(E_{(2)}\) denote the annual generation of the two remaining resource classes with the lowest levelised electricity costs. For the first step, set \(S_0=S^{\mathrm{base}}\). For subsequent steps, set \[ S^{\min} = \eta\left(E_{(1)}+0.2E_{(2)}\right), \qquad S_k = \max\left(S^{\min},S^{\mathrm{base}}\right). \]
\item \emph{Solve.} Solve Eqs.~\eqref{eq:si_s2_obj}--\eqref{eq:si_s2_ely_coupling} for the production target \(S_k\). If the problem is infeasible, the remaining potential is exhausted and the procedure stops.
\item \emph{Evaluate.} Calculate the year-specific total costs \(\mathrm{TC}_{y,k}\) using the optimal capacities. The corresponding levelised costs of hydrogen are \[ \mathrm{LCOH}_{y,k} = \frac{ \mathrm{TC}_{y,k} }{ \sum_{h\in\mathcal{H}}Q^{\star}_{y,h} } \qquad \forall\,y\in\mathcal{Y}. \]
\item \emph{Accept or refine.} Accept the first step \((k=0)\) automatically. For subsequent steps, accept the step if \[ \left| \mathrm{LCOH}_{2030,k} - \mathrm{LCOH}_{2030,k-1} \right| \leq\varepsilon, \] or if \(S_k=S^{\mathrm{base}}\). Otherwise, reject the step, set \[ S_k \leftarrow \max\left(0.25S_k,S^{\mathrm{base}}\right), \] and return to point~3.
\item \emph{Update.} After accepting the step, record \(S_k\), the year-specific values \(\mathrm{LCOH}_{y,k}\), and the optimal capacities for both support years. Deduct the deployed renewable capacity from the remaining potential: \[ \bar{K}_{y,t} \leftarrow \bar{K}_{y,t} - K^{g\star}_{y,t}. \] Set \(k\leftarrow k+1\) and return to point~2. \end{enumerate}

The procedure produces larger steps where costs change little and smaller steps where costs rise quickly. This avoids creating many small steps with nearly identical costs and limits the number of supply nodes used in the market model.

Renewable potentials differ between the support years, for instance because some offshore wind sites become available only in 2050. Once no further step can be constructed jointly for both support years, the procedure continues for 2050 alone, without the coupling constraints. The 2050 supply curve may therefore extend beyond the 2030 curve.

\subsection{Supply curve output and interfacing with the market model}

Each accepted step \(k\) becomes one supply node \(s\) in the market model. For each support year, the step's total investment and annual fixed O\&M costs are divided by its annual hydrogen output \(S_k\) to obtain the specific investment and fixed operating costs used in the market model: \begin{equation} c^{\mathrm{ely,inv}}_{y,s} = \frac{\text{total investment of step }k}{S_k}, \qquad c^{\mathrm{ely,fix}}_{y,s} = \frac{\text{annual fixed O\&M of step }k}{S_k}, \qquad \bar{G}_{y,s} = S_k. \label{eq:si_s2_output} \end{equation} The levelised cost of hydrogen for supply step \(s\), \[ \mathrm{LCOH}_{y,s} = \alpha^{\mathrm{ely}}_{y}c^{\mathrm{ely,inv}}_{y,s} + c^{\mathrm{ely,fix}}_{y,s}, \] is the LCOH assigned to a supply vintage in Eq.~\eqref{eq:si_lcoh_vintage}. Three post-processing operations are applied before the curves enter the market model:

\begin{enumerate}\setlength{\itemsep}{1pt}
\item \emph{Electrolysis cost.}
Electrolysis CAPEX is calculated by multiplying each step’s optimal electrolyser capacity by the year-specific CAPEX per unit of capacity. Fixed O\&M is calculated as a fixed share of this investment (Supplementary Table~S5). Renewable generation costs remain as determined by the supply-curve optimisation. This allows updated electrolyser CAPEX assumptions to be applied without rerunning the supply-curve optimisation, while retaining the optimised capacity composition of each step.

\item \emph{Steps present in one year only.}
Steps that exist in 2050 but
not in 2030 are assigned zero volume in 2030. To provide their costs for the intervening years, a 2030 reference cost is estimated from the cost ratios of steps available in both years. This reference cost is used only for interpolation and does not make the step available in 2030. The step becomes available gradually as its volume increases towards 2050.

\item \emph{Merging and interpolation.} Adjacent steps with identical
specific investment and fixed operating costs are combined and their volumes added. This reduces the number of supply nodes without changing the cost--volume relationship.
Costs and volumes are then linearly interpolated between the two support
years to obtain values for every model year from 2030 to 2050.
\end{enumerate}

The resulting curves differ markedly in shape across regions. Regions with large, high-yield resources of similar quality produce long and flat curves, whereas regions with smaller or more varied resource potentials produce short and steep ones.

\section{Sensitivity calculations}
\label{sec:si_sensitivity}
We test how the results change under different hydrogen-demand pathways. We repeat the global-market configuration and the single-buyer configurations for Germany and Japan under the Announced Pledges (APS) and Hydrogen Run-Up (HRU) scenarios from the StEAM dataset \cite{Linsel.2025,Linsel.2025b}. In 2045, global hydrogen demand is 36\% lower under APS and 47\% higher under HRU than in the base RES scenario. The same hydrogen supply curves are used in all three demand scenarios.

Relative to the base case, the demand-weighted price is 4\% lower in the APS scenario and 5\% higher in the HRU scenario, while the median regional price changes by no more than 8\%. Regional prices are averaged using the same base-case demand shares in all three scenarios. Changes are much larger in some regions and do not all move in the same direction. Under HRU, demand in Switzerland, Norway and IN-North increases by 47\%, while their prices change by \(+35\%\), \(+33\%\) and \(-11\%\), respectively. Regional prices therefore depend not only on their own demand, but also on how demand elsewhere changes which production and transport options remain available.

Germany's average delivered-cost benchmark changes by less than 1~EUR/MWh and Japan's by approximately 1~EUR/MWh. The price--cost gaps are 62, 71 and 91~EUR/MWh in Germany and 79, 96 and 97~EUR/MWh in Japan under APS, RES (Base Case) and HRU, respectively (Supplementary Table~S\ref{tab:si_demand_sensitivity}). The gap therefore remains substantial under all three demand pathways.

\clearpage
\section*{Supplementary Figures}
\addcontentsline{toc}{section}{Supplementary Figures}

\clearpage
\section*{Supplementary Tables}
\addcontentsline{toc}{section}{Supplementary Tables}

\begin{table}[!htbp]
\centering
\small
\caption{Country-level regions represented in HyPriM. Each of the 47
regions covers one country and is identified by its ISO~3166-1 alpha-2
code. A country-level model region may combine several of the 264 regions
used in \cite{Weienburger.2026}. Together with the 34 subnational and multi-country
regions listed in Supplementary Table~\ref{tab:si_regions_clusters},
these form the 81 model regions used in the present analysis.}
\label{tab:si_regions_countries}
\begin{tabular}{@{}ll@{\hspace{2em}}ll@{\hspace{2em}}ll@{}}
\toprule
Code & Country & Code & Country & Code & Country \\
\midrule
AE & United Arab Emirates & FR & France              & NZ & New Zealand \\
AR & Argentina            & GB & United Kingdom      & OM & Oman \\
AT & Austria              & GR & Greece              & PE & Peru \\
BE & Belgium              & HR & Croatia             & PL & Poland \\
BG & Bulgaria             & HU & Hungary             & PT & Portugal \\
CA & Canada               & ID & Indonesia           & RO & Romania \\
CH & Switzerland          & IE & Ireland             & SA & Saudi Arabia \\
CL & Chile                & IS & Iceland             & SE & Sweden \\
CZ & Czech Republic       & IT & Italy               & SI & Slovenia \\
DE & Germany              & JP & Japan               & SK & Slovakia \\
DK & Denmark              & KR & Republic of Korea   & TR & T\"urkiye \\
DZ & Algeria              & KZ & Kazakhstan          & TW & Taiwan \\
EG & Egypt                & MA & Morocco             & UA & Ukraine \\
ES & Spain                & MX & Mexico              & VN & Vietnam \\
FI & Finland              & NG & Nigeria             & ZA & South Africa \\
   &                      & NL & Netherlands         &    &              \\
   &                      & NO & Norway              &    &              \\
\bottomrule
\end{tabular}
\end{table}

\begingroup
\small
\begin{longtable}{@{}llp{0.49\textwidth}@{}}
\caption{Subnational and multi-country regions represented in HyPriM.
The remaining 34 model regions represent subnational areas, groups of countries, or combinations of the two. They are based on the regions used in \cite{Weienburger.2026}, some of which were aggregated for this analysis.}
\label{tab:si_regions_clusters}\\

\toprule
Model code & Region name & Geographic coverage \\
\midrule
\endfirsthead

\multicolumn{3}{@{}l}{\tablename~\thetable\ continued}\\
\addlinespace[0.3em]
\toprule
Model code & Region name & Geographic coverage \\
\midrule
\endhead

\midrule
\multicolumn{3}{r@{}}{Continued on next page}\\
\endfoot

\bottomrule
\endlastfoot

ASEANMainland  & ASEAN Mainland          & Cambodia, Laos, Myanmar, Thailand \\
ASEANMaritime  & ASEAN Maritime          & Brunei, Malaysia, Philippines, Singapore \\
AUEast         & Australia East          & New South Wales, South Australia, Tasmania, Victoria \\
AUWest         & Australia West          & Northern Territory, Queensland, Western Australia \\
Baltic         & Baltic states           & Estonia, Latvia, Lithuania \\
BRNorth        & Brazil North            & Brazil: Centre-North, North-East, North-West \\
BRSouth        & Brazil South            & Brazil: Centre-West, South, South-East, West \\
CAfrica        & Central Africa          & Cameroon, Central African Republic, Chad, Congo,
                                           Democratic Republic of the Congo, Equatorial Guinea,
                                           Gabon, S\~ao Tom\'e and Pr\'{\i}ncipe \\
Caucasus       & Caucasus                & Armenia, Azerbaijan, Georgia \\
CentralAm      & Central America and Carribean        & Bahamas, Belize, Costa Rica, Cuba, Dominican Republic,
                                           El Salvador, Guatemala, Haiti, Honduras, Jamaica,
                                           Nicaragua, Panama, Trinidad and Tobago \\
CentralAsia    & Central Asia            & Kyrgyzstan, Tajikistan, Turkmenistan, Uzbekistan \\
CNEast         & China East              & Anhui, Fujian, Jiangsu, Jiangxi, Shanghai, Zhejiang \\
CNNorthCoast   & China North Coast       & Beijing, Hebei, Liaoning, Shandong, Tianjin \\
CNNortheast    & China Northeast         & Heilongjiang, Inner Mongolia, Jilin, Shanxi \\
CNSouthCentral & China South-Central     & Chongqing, Guangdong, Guangxi, Guizhou, Hainan, Henan,
                                           Hubei, Hunan, Sichuan, Yunnan \\
CNWest         & China West and Mongolia             & Gansu, Ningxia, Qinghai, Shaanxi, Xinjiang, Xizang;
                                           Mongolia \\
EAfrica        & East Africa             & Burundi, Djibouti, Eritrea, Ethiopia, Kenya, Rwanda,
                                           Somalia, South Sudan, Sudan, Tanzania, Uganda \\
GulfNorth      & Gulf region, other           & Bahrain, Iran, Iraq, Kuwait, Qatar, Yemen \\
INNorth        & India North             & India: East, North, North-East \\
INSouth        & India South             & India: South, West \\
Levant         & Levant                  & Israel, Jordan, Lebanon, Palestine, Syria \\
LibyaTunisia   & Libya and Tunisia       & Libya, Tunisia \\
NAMBot         & Namibia and Botswana    & Botswana, Namibia \\
RUEast         & Russia East             & Russia: Far East, Siberia \\
RUWest         & Russia West             & Russia: Centre, Middle Volga, North-West, South, Ural;
                                           Belarus \\
SAfrOther      & Southern Africa, other  & Angola, Eswatini, Lesotho, Madagascar, Malawi,
                                           Mauritius, Mozambique, Zambia, Zimbabwe \\
SAMOther       & South America, other    & Bolivia, Colombia, Ecuador, Guyana, Paraguay,
                                           Suriname, Uruguay, Venezuela \\
SouthAsia      & South Asia              & Afghanistan, Bangladesh, Bhutan, Nepal, Pakistan,
                                           Sri Lanka \\
USCentral      & United States Central   & United States: Mid-North, Midwest North, Midwest South \\
USEast         & United States East      & United States: Mid-East North, Mid-East South, Northeast \\
USSouth        & United States South     & United States: Mid-South, South-East \\
USWest         & United States West      & United States: North-West, South-West \\
WAfrica        & West Africa             & Benin, Burkina Faso, Cabo Verde, C\^ote d'Ivoire, Gambia,
                                           Ghana, Guinea, Guinea-Bissau, Liberia, Mali, Mauritania,
                                           Niger, Senegal, Sierra Leone, Togo \\
WBalkans       & Western Balkans         & Albania, Bosnia and Herzegovina, Kosovo, Montenegro,
                                           North Macedonia, Serbia \\

\end{longtable}
\endgroup

\begin{table}[!htbp]
\centering
\small
\caption{Meta data used to obtain the country-level WACCs from the
Global WACC Calculator \cite{Jansen_Global_WACC_Calculator_2026}. Country-specific equity risk
premia, country risk premia and corporate tax rates are supplied by the
calculator's underlying datasets. The reported model inputs are real WACCs.}
\label{tab:si_wacc_method}
\begin{tabular}{@{}ll@{}}
\toprule
Setting & Value \\
\midrule
Risk-free rate & 3.5\% \\
Unlevered beta & 1.10 \\
Country-specific equity risk premium & Enabled \\
Swap rate & 3.0\% \\
Credit margin & 2.0\% \\
Cost of debt & 5.0\% \\
Equity / debt share & 40\% / 60\% \\
Inflation rate for conversion to real WACC & 2.0\% \\
\bottomrule
\end{tabular}
\end{table}

\begin{table}[!htbp]
\centering
\small
\caption{Real weighted average cost of capital used for each of the
81 HyPriM model regions. Values are shown to one decimal place.
Supplementary Table~\ref{tab:si_wacc_method} gives the meta data. Model codes are ordered alphabetically down each column.}
\label{tab:si_wacc_regions}
\begin{tabular}{@{}lr@{\hspace{2em}}lr@{\hspace{2em}}lr@{}}
\toprule
Region & WACC (\%) & Region & WACC (\%) & Region & WACC (\%) \\
\midrule
AE & 6.5 & DZ & 11.7 & NO & 5.8 \\
AR & 24.9 & EAfrica & 20.5 & NZ & 5.4 \\
ASEANMainland & 17.3 & EG & 19.0 & OM & 10.7 \\
ASEANMaritime & 7.6 & ES & 8.4 & PE & 8.1 \\
AT & 6.3 & FI & 6.6 & PL & 7.5 \\
AUEast & 5.3 & FR & 6.6 & PT & 8.0 \\
AUWest & 5.3 & GB & 6.6 & RO & 10.1 \\
Baltic & 7.6 & GR & 10.2 & RUEast & 11.3 \\
BE & 6.6 & GulfNorth & 16.2 & RUWest & 15.7 \\
BG & 9.4 & HR & 8.2 & SA & 6.9 \\
BRNorth & 9.2 & HU & 10.1 & SAfrOther & 17.0 \\
BRSouth & 9.2 & ID & 9.1 & SAMOther & 16.9 \\
CA & 5.5 & IE & 7.4 & SE & 5.8 \\
CAfrica & 19.3 & INNorth & 9.0 & SI & 8.1 \\
Caucasus & 11.5 & INSouth & 9.0 & SK & 8.0 \\
CentralAm & 15.5 & IS & 7.1 & SouthAsia & 16.1 \\
CentralAsia & 15.6 & IT & 9.5 & TR & 13.5 \\
CH & 6.2 & JP & 6.5 & TW & 6.9 \\
CL & 6.9 & KR & 6.5 & UA & 27.7 \\
CNEast & 6.8 & KZ & 8.7 & USCentral & 5.6 \\
CNNorthCoast & 6.8 & Levant & 27.2 & USEast & 5.6 \\
CNNortheast & 6.8 & LibyaTunisia & 15.8 & USSouth & 5.6 \\
CNSouthCentral & 6.8 & MA & 9.4 & USWest & 5.6 \\
CNWest & 8.0 & MX & 8.5 & VN & 11.3 \\
CZ & 7.0 & NAMBot & 10.3 & WAfrica & 18.4 \\
DE & 5.3 & NG & 18.1 & WBalkans & 14.1 \\
DK & 5.8 & NL & 5.5 & ZA & 10.7 \\
\bottomrule
\end{tabular}
\end{table}

\begin{table}[!htbp]
\centering
\small
\setlength{\tabcolsep}{4pt}
\caption{Techno-economic assumptions of the market model. Monetary values
are in 2025 euros. Values without year labels apply throughout the modelled
period. For time-dependent inputs, selected years are shown. Values for
intervening model years are interpolated between the underlying data points.
Pipeline values are ordered by size class S, M and L. Pipeline investment
costs are base values before application of the route-specific cost factors
\cite{Weienburger.2026}. The electricity price for transport and conversion
is uniform across regions. Renewable-generation costs enter through the
supply curves (Supplementary Note~\ref{sec:si_supply}).}
\label{tab:si_tech_inputs}
\begin{tabular}{@{}p{0.25\textwidth}p{0.18\textwidth}p{0.35\textwidth}p{0.17\textwidth}@{}}
\toprule
Parameter & Unit & Value & Source \\
\midrule
\multicolumn{4}{@{}l}{\itshape Electrolysis} \\
Investment cost & EUR/kW & 2030: 2084, 2040: 1415, 2050: 1158
 & \cite{Klaas.2025} \\
Annual fixed O\&M & \% of CAPEX & 2 & \cite{Klaas.2025} \\
Lifetime & years & 25 & \cite{Klaas.2025} \\
Efficiency & \%
 & 2030: 69, 2035: 70, 2040: 72, 2050: 74
 & \cite{Klaas.2025} \\
Maximum annual increase in hydrogen production per region
 & TWh/year & 2030: 15, 2040: 156, 2050: 296
 & own assumption, 2030 based on \cite{InternationalEnergyAgencyProjects.2025}\\
\addlinespace
\multicolumn{4}{@{}l}{\itshape Pipeline (S / M / L)} \\
Investment cost & MEUR/km & 1.46 / 2.41 / 3.61
 & \cite{EuropeanHydrogenBackboneInitiative.2022} \\
Capacity & TWh/year & 13.5 / 30.4 / 54.1
 & \cite{EuropeanHydrogenBackboneInitiative.2022} \\
Annual fixed O\&M & \% of CAPEX & 5 & \cite{Weienburger.2026} \\
Lifetime & years & 40 & \cite{Weienburger.2026} \\
Compression electricity per 1000 km
 & MWh$_{\mathrm{el}}$/MWh$_{\mathrm{H_2}}$
 & 0.09 & Own assumption \\
\addlinespace
\multicolumn{4}{@{}l}{\itshape Liquefaction terminal} \\
Investment cost & MEUR/(TWh/year)
 & 2030: 241, 2040: 208, 2050: 180 & \cite{Klaas.2025} \\
Annual fixed O\&M & \% of CAPEX & 4 & \cite{Klaas.2025} \\
Lifetime & years & 30 & \cite{Klaas.2025} \\
Electricity consumption & MWh$_{\mathrm{el}}$/MWh$_{\mathrm{H_2}}$
 & 0.22 & \cite{Klaas.2025} \\
\addlinespace
\multicolumn{4}{@{}l}{\itshape Regasification terminal} \\
Investment cost & MEUR/(TWh/year) & 16.3 & \cite{Klaas.2025} \\
Annual fixed O\&M & \% of CAPEX & 4 & \cite{Klaas.2025} \\
Lifetime & years & 30 & \cite{Klaas.2025} \\
Electricity consumption & MWh$_{\mathrm{el}}$/MWh$_{\mathrm{H_2}}$
 & 0.016 & \cite{Klaas.2025} \\
\addlinespace
\multicolumn{4}{@{}l}{\itshape LH$_2$ carrier} \\
Investment cost & MEUR/vessel & 584 & \cite{Klaas.2025} \\
Cargo capacity & TWh/vessel & 0.290 & \cite{Klaas.2025} \\
Annual fixed O\&M & \% of CAPEX & 4 & \cite{Klaas.2025} \\
Variable cost & MEUR/voyage & 0.301
 & Own assumption \\
Lifetime & years & 15 & \cite{Klaas.2025} \\
Speed & km/h & 30 & \cite{Klaas.2025} \\
Port time per round trip & hours & 150 & Own assumption \\
Boil-off during one-way voyage & \%/day & 0.6 & \cite{Klaas.2025} \\
H$_2$ consumption per km travelled on both legs
 & MWh/km & 0.41 & \cite{EWI.GlobalH2CostToolV3} \\
\addlinespace
\multicolumn{4}{@{}l}{\itshape Other} \\
Electricity price for transport and conversion & EUR/MWh
 & 2030: 45.1, 2040: 42.1, 2050: 39.1 & \cite{Klaas.2025} \\
\bottomrule
\end{tabular}
\end{table}

\begin{table}[!htbp]
\centering
\caption{Sets, indices and mappings of the HyPriM market model. The model
uses annual time steps from 2030 to 2050. The index \(v\) identifies the
year in which an asset is commissioned.}
\label{tab:si_sets}
\begin{tabular}{@{}ll@{}}
\toprule
Symbol & Description \\
\midrule
\multicolumn{2}{@{}l}{\itshape Indices and index sets} \\
\addlinespace[0.15em]
\(y \in Y\) & Model year, \(Y = \{2030,\dots,2050\}\);
\(y_0 = \min Y\), \(\bar{y} = \max Y\) \\
\(v \in Y\) & Commissioning year (vintage), with \(v \le y\) \\
\(s \in S\) & Supply node; each \(s\) is one step of a regional supply curve \\
\(h \in H\) & Hub (modelled region) \\
\(p \in P\) & Port \\
\(d \in D\) & Hydrogen demand node \\
\(z \in Z\) & Pipeline size class,
\(Z = \{\mathrm{S},\mathrm{M},\mathrm{L}\}\) for small, medium and large \\
\addlinespace
\multicolumn{2}{@{}l}{\itshape Mappings} \\
\addlinespace[0.15em]
\(S(h) \subseteq S\) & Supply nodes located at hub \(h\) \\
\(h(p)\) & Hub associated with port \(p\) \\
\(h(d)\) & Hub associated with demand node \(d\) \\
\bottomrule
\end{tabular}
\end{table}

\begin{table}[!htbp]
\centering
\caption{Arc sets of the HyPriM market model. For any arc \(a\), \(o(a)\) and \(e(a)\) denote its origin and destination, respectively. All arc types except pipelines are directed. Each connectable pair of hubs is represented by one pipeline arc. Its listed direction defines a positive flow, while a negative flow represents transport in the opposite direction.}
\label{tab:si_arcs}
\begin{tabular}{@{}ll@{}}
\toprule
Symbol & Description \\
\midrule
\(\mathcal{A}^{\mathrm{sh}} \subseteq S \times H\)
& Supply node to regional hub \\
\(\mathcal{A}^{\mathrm{pipe}} \subseteq H \times H\)
& Pipeline connection between hubs \\
\(\mathcal{A}^{\mathrm{hp}} \subseteq H \times P\)
& Hub to port for liquefaction \\
\(\mathcal{A}^{\mathrm{ph}} \subseteq P \times H\)
& Port to hub after regasification \\
\(\mathcal{A}^{\mathrm{ship}} \subseteq P \times P\)
& Directed shipping route between ports \\
\(\mathcal{A}^{\mathrm{hd}} \subseteq H \times D\)
& Hub to hydrogen demand node \\
\bottomrule
\end{tabular}
\end{table}

\begin{table}[!htbp]
\centering
\caption{Decision variables of the HyPriM market model. Investment variables record additions in year \(y\). Available capacities and the number of pipelines are obtained by accumulating these additions over vintages, as defined in Eq.~\eqref{eq:si_vintage}.}
\label{tab:si_vars}
\begin{tabularx}{\textwidth}{@{}ll>{\raggedright\arraybackslash}X@{}}
\toprule
Symbol & Domain & Description \\
\midrule
\multicolumn{3}{@{}l}{\itshape Operational variables
\textnormal{(optimised in both stages)}} \\
\addlinespace[0.15em]
\(g_{y,s}\)
& \(\mathbb{R}_+\)
& Hydrogen produced at supply node \(s\) (TWh/a) \\
\(l_{y,p}\)
& \(\mathbb{R}_+\)
& Hydrogen liquefied at port \(p\) (TWh/a) \\
\(r_{y,p}\)
& \(\mathbb{R}_+\)
& Hydrogen regasified at port \(p\) (TWh/a) \\
\(f_{y,a}\)
& \(\mathbb{R}_+\)
& Flow on arc \(a \notin \mathcal{A}^{\mathrm{pipe}}\) (TWh/a) \\
\(f_{y,a}\)
& \(\mathbb{R}\)
& Signed flow on \(a \in \mathcal{A}^{\mathrm{pipe}}\) (TWh/a) \\
\(\tilde{f}_{y,a}\)
& \(\mathbb{R}_+\)
& Absolute pipeline flow, \(a \in \mathcal{A}^{\mathrm{pipe}}\) (TWh/a) \\
\(u_{y,d}\)
& \(\mathbb{R}_+\)
& Unserved demand at node \(d\) (TWh/a) \\
\addlinespace
\multicolumn{3}{@{}l}{\itshape Investment variables
\textnormal{(optimised in Stage~1 and fixed in Stage~2)}} \\
\addlinespace[0.15em]
\(K^{\mathrm{ely}}_{y,s}\)
& \(\mathbb{R}_+\)
& Hydrogen-production capacity added at supply node \(s\), expressed as annual output (TWh/a) \\
\(K^{\mathrm{liq}}_{y,p}\)
& \(\mathbb{R}_+\)
& Liquefaction capacity added (TWh/a) \\
\(K^{\mathrm{reg}}_{y,p}\)
& \(\mathbb{R}_+\)
& Regasification capacity added (TWh/a) \\
\(K^{\mathrm{shp}}_{y}\)
& \(\mathbb{R}_+\)
& Cargo capacity added to the global shipping fleet (TWh) \\
\(n_{y,a,z}\)
& \(\mathbb{Z}_+\)
& Number of pipelines of size \(z\) added on arc \(a\) \\
\bottomrule
\end{tabularx}
\end{table}

\begin{table}[!htbp]
\centering
\caption{Parameters of the HyPriM market model. Annual flows and throughput
capacities are expressed in TWh/a, while shipping cargo capacity is
expressed in TWh. Monetary values are given in 2025 euros. Costs per unit
of throughput are reported in EUR/MWh, which is numerically identical to
MEUR/TWh when combined with quantities in TWh. The superscript
\(\theta\) indexes technologies, with
\(\theta \in
\{\mathrm{ely},\mathrm{pipe},\mathrm{liq},\mathrm{reg},\mathrm{shp}\}\).}
\label{tab:si_params}
\begin{tabularx}{\textwidth}{@{}l>{\raggedright\arraybackslash}X@{}}
\toprule
Symbol & Description \\
\midrule

\multicolumn{2}{@{}l}{\itshape Demand} \\
\addlinespace[0.15em]
\(q_{y,d}\)
& Exogenous hydrogen demand at node \(d\) (TWh/a) \\
\(\varphi_{y,d}\)
& Penalty on unserved demand, set to 999~EUR/MWh \\
\(\mu\)
& Stage~1 adequacy multiplier, \(\mu=1.05\), corresponding to a 5\% margin \\
\addlinespace

\multicolumn{2}{@{}l}{\itshape Hydrogen supply} \\
\addlinespace[0.15em]
\(\bar{G}_{y,s}\)
& Maximum installed hydrogen-production capacity at supply node \(s\) (TWh/a) \\
\(\bar{\Delta}_{y,h}\)
& Maximum increase in hydrogen production between consecutive model years at hub \(h\) (TWh/a) \\
\addlinespace

\multicolumn{2}{@{}l}{\itshape Pipeline network} \\
\addlinespace[0.15em]
\(\kappa^{\mathrm{pipe}}_{y,z}\)
& Throughput capacity of one pipeline of size \(z\) (TWh/a) \\
\(\bar{n}\)
& Maximum number of pipelines added per arc and year, \(\bar{n}=3\) \\
\addlinespace

\multicolumn{2}{@{}l}{\itshape Shipping} \\
\addlinespace[0.15em]
\(\delta_a\)
& Routing distance of shipping route \(a\) (km) \\
\(\nu_y\)
& Ship speed (km/h) \\
\(\tau^{\mathrm{port}}\)
& Fixed port time per round trip, \(\tau^{\mathrm{port}}=150\)~h \\
\(\tau_{y,a}=2\delta_a/\nu_y+\tau^{\mathrm{port}}\)
& Round-trip time on shipping route \(a\) (h) \\
\(\ell_{y,a}\in[0,1)\)
& Shipping losses from boil-off and vessel self-consumption \\
\(\kappa^{\mathrm{shp}}_y\)
& Cargo capacity of one vessel (TWh) \\
\addlinespace

\multicolumn{2}{@{}l}{\itshape Cost of capital and lifetimes} \\
\addlinespace[0.15em]
\(L^\theta_y\)
& Technical lifetime of technology \(\theta\) (a) \\
\(\rho_h\)
& Weighted average cost of capital of region \(h\) \\
\(\alpha^\theta_y\)
& Annuity factor calculated from the applicable cost of capital and technical lifetime \\
\(\omega^\theta_y
 = \min(\bar{y}-y+1,L^\theta_y)\)
& Number of model years for which a vintage-\(y\) asset is charged \\
\addlinespace

\multicolumn{2}{@{}l}{\itshape Costs} \\
\addlinespace[0.15em]
\(c^{\theta,\mathrm{inv}}_{y,\cdot}\)
& Specific investment cost per unit of the corresponding capacity variable \\
\(c^{\theta,\mathrm{fix}}_{y,\cdot}\)
& Annual fixed O\&M cost per unit of the corresponding capacity variable \\
\(c^{\theta,\mathrm{var}}_{y,\cdot}\)
& Variable cost per unit of throughput (EUR/MWh) \\
\(c^{\mathrm{el}}_y\)
& Electricity price for compression, liquefaction and regasification (EUR/MWh) \\
\(e^{\mathrm{pipe}}_y\)
& Electricity consumption of pipeline transport per unit of hydrogen and kilometre \\
\(e^{\mathrm{liq}}_y,\ e^{\mathrm{reg}}_y\)
& Electricity consumption per unit of hydrogen liquefied or regasified \\
\bottomrule
\end{tabularx}
\end{table}

\begin{table}[!htbp]
\centering
\caption{Sets, indices and parameters of the hydrogen supply-curve model.
The notation is local to this formulation. In particular, \(h\) denotes an
hour, whereas it denotes a hub in Supplementary
Note~\ref{sec:si_hyprim}. The model is solved separately for each region,
using the same regional weighted average cost of capital \(\rho\) for
renewable generation and electrolysis. Monetary values are given in 2025
euros.}
\label{tab:si_s2_params}
\begin{tabularx}{\textwidth}{@{}l>{\raggedright\arraybackslash}X@{}}
\toprule
Symbol & Description \\
\midrule

\multicolumn{2}{@{}l}{\itshape Sets and indices} \\
\addlinespace[0.15em]
\(y \in \mathcal{Y}\)
& Support year, \(\mathcal{Y}=\{2030,2050\}\) \\
\(t \in \mathcal{T}\)
& Renewable resource class, defined by technology and resource quality. \\
&
Technologies include open-field PV, concentrating solar power, onshore wind
and offshore wind \\
\(h \in \mathcal{H}\)
& Hour of a representative year, \(|\mathcal{H}|=8760\) \\
\(k=0,1,2,\dots\)
& Supply-curve step \\
\addlinespace

\multicolumn{2}{@{}l}{\itshape Renewable resource data} \\
\addlinespace[0.15em]
\(\bar{K}^{\mathrm{raw}}_{y,t}\)
& Maximum installable capacity of resource class \(t\) (MW) \\
\(F_{y,t}\)
& Full-load hours of resource class \(t\) (h/a) \\
\(\phi_{y,h,t}\)
& Normalised hourly generation profile, with
\(\sum_{h\in\mathcal{H}}\phi_{y,h,t}=1\) \\
&
Available generation in hour \(h\) is
\(K^g_{y,t}\phi_{y,h,t}F_{y,t}\) (MWh) \\
\addlinespace

\multicolumn{2}{@{}l}{\itshape Costs and cost of capital} \\
\addlinespace[0.15em]
\(I^g_{y,t}\)
& Investment cost of renewable resource class \(t\) (EUR/MW) \\
\(f^g_{y,t}\)
& Annual fixed O\&M cost of renewable resource class \(t\)
(EUR/(MW\,a)) \\
\(L^g_{y,t}\)
& Technical lifetime of renewable resource class \(t\) (a) \\
\(I^e_y\)
& Electrolyser investment cost (EUR/MW\(_{\mathrm{H_2}}\)) \\
\(f^e_y\)
& Annual fixed electrolysis O\&M cost
(EUR/(MW\(_{\mathrm{H_2}}\)\,a)) \\
\(L^e_y\)
& Electrolyser lifetime (a) \\
\(\rho\)
& Regional weighted average cost of capital \\
\(\alpha^g_{y,t},\ \alpha^e_y\)
& Annuity factors derived from \(\rho\) and the respective lifetimes \\
\addlinespace

\multicolumn{2}{@{}l}{\itshape Electrolysis and step construction} \\
\addlinespace[0.15em]
\(\eta\)
& Electrolyser efficiency
(MWh\(_{\mathrm{H_2}}\)/MWh\(_{\mathrm{el}}\)) \\
\(D^{\mathrm{el}}\)
& Domestic electricity demand to be reserved (MWh/a) \\
\(S^{\mathrm{base}}\)
& Minimum supply-curve step size (MWh\(_{\mathrm{H_2}}\)/a) \\
\(\varepsilon\)
& Threshold for refining the step size based on the change in 2030 LCOH
between adjacent steps (EUR/MWh) \\
\bottomrule
\end{tabularx}
\end{table}

\begin{table}[!htbp]
\centering
\caption{Decision variables of the hydrogen supply-curve model. The
optimisation is solved separately for each region and supply-curve step
\(k\), so the region and step indices are omitted. Electrolyser capacity is
expressed on the hydrogen-output side: \(K^e_y\) limits hourly hydrogen
production rather than electricity input.}
\label{tab:si_s2_vars}
\begin{tabular}{@{}lll@{}}
\toprule
Symbol & Domain & Description \\
\midrule

\multicolumn{3}{@{}l}{\itshape Capacity variables} \\
\addlinespace[0.15em]
\(K^g_{y,t}\)
& \(\mathbb{R}_+\)
& Renewable-generation capacity assigned to resource class \(t\) (MW) \\
\(K^e_y\)
& \(\mathbb{R}_+\)
& Electrolyser capacity expressed as hydrogen output
(MW\(_{\mathrm{H_2}}\)) \\
\addlinespace

\multicolumn{3}{@{}l}{\itshape Hourly operation} \\
\addlinespace[0.15em]
\(g_{y,h,t}\)
& \(\mathbb{R}_+\)
& Renewable electricity supplied to electrolysis by resource class \(t\)
(MWh) \\
\(G_{y,h}\)
& \(\mathbb{R}_+\)
& Total electricity input to electrolysis (MWh) \\
\(Q_{y,h}\)
& \(\mathbb{R}_+\)
& Hydrogen output (MWh\(_{\mathrm{H_2}}\)) \\
\bottomrule
\end{tabular}
\end{table}

\begin{table}[!htbp]
\centering
\small
\setlength{\tabcolsep}{4pt}
\caption{Demand sensitivity results for Germany and Japan in 2045. APS,
RES and HRU denote the Announced Pledges, Renewable Electrification and
Hydrogen Run-Up scenarios from the StEAM dataset
\cite{Linsel.2025,Linsel.2025b}. \(c_1\) and \(p_1\) denote the average
delivered cost and regional hydrogen price in the single-buyer
configuration, while \(c_N\) and \(p_N\) denote the corresponding quantities
in the global-market configuration. Components are calculated before
rounding and may therefore not sum exactly to the displayed totals.}
\label{tab:si_demand_sensitivity}
\begin{tabular}{@{}lrrrrrr@{}}
\toprule
& \multicolumn{3}{c}{Germany}
& \multicolumn{3}{c}{Japan} \\
\cmidrule(lr){2-4}\cmidrule(lr){5-7}
& APS & RES & HRU & APS & RES & HRU \\
\midrule

Regional hydrogen demand (TWh/a)
& 211 & 318 & 466 & 257 & 595 & 880 \\
\addlinespace

\multicolumn{7}{@{}l}{\itshape Costs and prices (EUR/MWh)} \\
\addlinespace[0.15em]
\(c_1\)
& 118.6 & 118.6 & 119.4 & 123.6 & 122.7 & 122.5 \\
\(p_1\)
& 123.8 & 128.9 & 127.2 & 123.6 & 150.1 & 159.2 \\
\(c_N\)
& 135.3 & 146.4 & 148.1 & 151.6 & 139.5 & 155.8 \\
\(p_N\)
& 180.5 & 189.2 & 210.2 & 202.6 & 219.1 & 219.5 \\
\addlinespace

\multicolumn{7}{@{}l}{\itshape Price--cost gap and decomposition (EUR/MWh)} \\
\addlinespace[0.15em]
\(\Gamma=p_N-c_1\)
& 61.9 & 70.6 & 90.8 & 79.0 & 96.5 & 96.9 \\
\(\Gamma\) as a share of \(c_1\) (\%)
& 52 & 60 & 76 & 64 & 79 & 79 \\
\(\Delta c=c_N-c_1\)
& 16.7 & 27.8 & 28.7 & 28.1 & 16.9 & 33.2 \\
\(\Delta^{pc}_{N}=p_N-c_N\)
& 45.2 & 42.8 & 62.0 & 51.0 & 79.6 & 63.7 \\
\(\Delta^{pc}_{1}=p_1-c_1\)
& 5.2 & 10.3 & 7.8 & 0.0 & 27.5 & 36.6 \\
\(\Delta p=p_N-p_1\)
& 56.7 & 60.3 & 83.0 & 79.0 & 69.0 & 60.3 \\
\bottomrule
\end{tabular}
\end{table}

\end{document}